\documentclass[12pt,a4paper]{article}
\usepackage[top=2.25cm,bottom=2.25cm,left=2.05cm,right=2.05cm]{geometry}
\usepackage{latexsym,amsmath,amsfonts,amssymb,amsthm,amscd,mathrsfs}

\usepackage[usenames,dvipsnames]{color}
\usepackage{tikz}
\usepackage{tikz-cd}
\usetikzlibrary{matrix,arrows,decorations.pathmorphing}

\newcommand{\be}{\begin{equation}}
\newcommand{\ee}{\end{equation}}
\newcommand{\bea}{\begin{eqnarray}}
\newcommand{\eea}{\end{eqnarray}}
\newcommand{\ba}{\begin{aligned}}
\newcommand{\ea}{\end{aligned}}
\newcommand{\bpm}{\begin{pmatrix}}
\newcommand{\epm}{\end{pmatrix}}

\numberwithin{equation}{section}
\newcounter{thmcounter}
\numberwithin{thmcounter}{section}

\theoremstyle{definition}

\theoremstyle{plain}

\newtheorem{lemma}[thmcounter]{Lemma}
\newtheorem{proposition}[thmcounter]{Proposition}

\newenvironment{proofof}[2]{{\itshape \noindent Proof of #1 \ref{#2}}}{\hfill\(\square\)}

\DeclareMathOperator*{\rez}{Res}

\def\cB{{\mathcal B}}                       %
\def\cF{{\mathcal F}}                       %
\def\cH{{\mathcal H}}                       %
\def\cK{{\mathcal K}}                       %
\def\cL{{\mathcal L}}                       %
\def\cM{{\mathcal M}}                       %
\def\cN{{\mathcal N}}                       %
\def\cP{{\mathcal P}}                       %
\def\cR{{\mathcal R}}                       %
\def\cS{{\mathcal S}}                       %
\def\cW{{\mathcal W}}                       %
\def\cZ{{\mathcal Z}}                       %
\def\cD {\mathscr{D}}                       %
\def\C{\mathbb{C}}                          %
\def\D{\mathbb{D}}                          %
\def\R{\mathbb{R}}                          %
\def\T{\mathbb{T}}                          %
\def\Z{\mathbb{Z}}                          %
\def\fH{\mathfrak{H}}                       %
\def\fP{\mathfrak{P}}                       %
\def\diag{\mathrm{diag}}                    %
\def\ri{{\rm i}}                            %
\def\tr{\mathrm{tr}}                        %
\def\BC{\mathrm{BC}}                        %
\def\GL{{\rm GL}}                           %
\def\SL{{\rm SL}}                           %
\def\SB{{\rm SB}}                           %
\def\U{{\rm U}}                             %
\def\SU{{\rm SU}}                           %
\def\sl{{\rm sl}}                           %
\def\can{\mathrm{can}}                      %
\def\red{\mathrm{red}}                      %
\def\reg{\mathrm{reg}}                      %
\def\sreg{{\rm sreg}}                       %
\def\1{{\boldsymbol 1}}                     %
\def\0{{\boldsymbol 0}}                     %
\def\ds{\left.\frac{d}{ds}\right\vert_{s=0}}%

\begin{document}
\begin{center}
{\large\bf
Global description of action-angle duality for a Poisson-Lie deformation of the
 trigonometric $\boldsymbol{\BC_n}$ Sutherland system}
\end{center}

\smallskip
\begin{center}
L.~Feh\'er${}^{a,b}$ and I.~Marshall${}^c$\\

\bigskip
${}^a$Department of Theoretical Physics, University of Szeged\\
Tisza Lajos krt 84-86, H-6720 Szeged, Hungary\\
e-mail:  lfeher@physx.u-szeged.hu

\smallskip
${}^b$Department of Theoretical Physics, WIGNER RCP, RMKI\\
H-1525 Budapest, P.O.B.~49, Hungary\\

\medskip
${}^c$
Faculty of Mathematics, Higher School of Economics\\
National Research University\\
Usacheva 6, Moscow, Russia\\
 e-mail: imarshall@hse.ru

\end{center}

\smallskip
\begin{abstract}
Integrable many-body systems of Ruijsenaars--Schneider--van Diejen
type displaying action-angle duality are derived  by
Hamiltonian reduction of the Heisenberg double of the Poisson-Lie
group $\SU(2n)$.
New global models of the reduced phase space are described, revealing
non-trivial
features of the two systems in duality with one another.
For example, after establishing that the symplectic vector space $\C^n\simeq\R^{2n}$ underlies
both global models, it is seen that for both systems the action variables
generate the standard torus action on $\C^n$, and the fixed point of this action corresponds
to the unique equilibrium positions of the pertinent
systems.
The systems in duality are found to be  non-degenerate in the sense
that the functional dimension of the Poisson algebra of their
conserved quantities
is equal to half the dimension of the phase space.
The dual of the deformed Sutherland system is shown to be a limiting
case of a van~Diejen system.

\end{abstract}

{\linespread{0.8}\tableofcontents}

\newpage
\section{Introduction}
\label{sec:1}

Integrable Hamiltonian systems have important applications in diverse fields of physics and are
in the focus of intense investigation by a great variety of mathematical methods.
We are interested in the family of classical many-body systems introduced in their
simplest form  by Calogero \cite{Cal}, Sutherland \cite{Suth} and Ruijsenaars and Schneider \cite{RS86}.
 The relevance of these systems
to numerous
areas of mathematics and physics
is apparent from the
reviews devoted to them \cite{vDV, EtiR, N, OP1, PolR, RuijKup, Banff, SuthR}.
One of their fascinating features is that several pairs of such systems enjoy a duality
relation
that converts the particle positions of one system into the
action variables of the other system, and vice versa\footnote{Self-duality occurs when the related systems are identical,
except for a possible shift of their parameters.}.
This intriguing phenomenon was first analyzed in the ground-breaking papers \cite{SR88,RIMS95} by a direct method,
while its group-theoretic background came to light more recently \cite{JHEP, G, N}.
The treatment of the self-dual Calogero system by Kazhdan, Kostant and Sternberg  \cite{KKS}
served as a source of inspiration for these developments.
Since this paper is devoted to the analysis of  a particular dual pair, let us next outline
in more precise terms the notion of duality that we use.

An integrable Hamiltonian system is given by an
Abelian Poisson algebra $\fH$ of smooth
functions on a $2n$-dimensional symplectic manifold
$(M,\omega)$ such that the functional dimension of $\fH$ is $n$, and all elements of $\fH$ generate  complete flows.
The systems of our interest possess another distinguished Abelian Poisson algebra $\fP$, which has the same
properties as $\fH$ and the following requirements hold:\\
(a) There exist Darboux coordinates, $\lambda_i, \theta_j$, on a dense open submanifold $M^o$ of $M$  such that
the restriction of $\fP$ to $M^o$ is functionally generated by  the $\lambda_i$.\\
(b) $\fH$ contains a distinguished function $H$ whose restriction to $M^o$ admits
interpretation as a many-body Hamiltonian describing the dynamics
of $n$ interacting `point-particles' with positions $\lambda_i$ moving along one dimensional space
(a line or a circle).\\
The function $H$ is often called the `main Hamiltonian' and
$\fP$ is sometimes called the algebra of  `global position variables'.

Now, suppose that we have two systems
\be
(M, \omega, \fH, \fP, H) \quad\hbox{and}\quad (\hat M, \hat\omega,  \hat\fH, \hat\fP, \hat H),
\label{I1}\ee
with associated Darboux coordinates, according to conditions (a) and (b), $(\lambda,\theta)$ and $(\hat\lambda,\hat\theta)$.
We say that these two systems are in action-angle duality (also called Ruijsenaars duality)
if there exists a \emph{global} symplectomorphism $\cR\colon (M, \omega)\to (\hat M,\hat\omega)$ such that
\be
\fH = \hat\fP\circ \cR
\quad\hbox{and}\quad
\hat\fH = \fP \circ \cR^{-1}.
\label{I2}\ee
An additional feature,
valid in all known examples,  is that the
Hamiltonian flows of $(M, \omega, \fP)$ and $(\hat M,  \hat\omega, \hat\fP)$
can be written down explicitly, not only on the dense open parts, but globally.
Consequently, $(M, \omega, \fH)$ is integrated by means of $(\hat M, \hat\omega, \hat\fP)$,
and $(\hat M, \hat\omega, \hat\fH)$ is integrated by means of $(M, \omega, \fP)$.
This means that $\cR$ and  $\cR^{-1}$ can be interpreted as global action-angle maps
for the Liouville integrable systems $(M, \omega,\fH)$ and $(\hat M, \hat\omega, \hat\fH)$.
One may also say that $\hat\fP$ represents global position-type variables
for the many-body system $(\hat M,  \hat\omega, \hat H)$ and
global action-type variables for the system $(M, \omega, H)$, together
with the analogous  `dual statement'.

For further description of this curious notion and its quantum mechanical counterpart,
alias the celebrated bispectral property \cite{DG}, the reader may consult the reviews \cite{RuijKup,Banff}.
We note in passing that in some examples the $\lambda_i$ are globally smooth and independent,
and then $M^o=M$, while
in other examples they lose their smoothness or independence outside a proper
submanifold $M^o$.
This should not come as a surprise since from the dual viewpoint the $\lambda_i$ are action variables,
which usually exhibit some singularities.
 Their canonical conjugates $\theta_i$ may vary on the circle or on the line
depending on the example.

It was realized by Gorsky and his collaborators \cite{JHEP,G,N}, and
explored  in detail by others (\cite{F-PLA}--\cite{FM}, \cite{P1,P}),  that dual pairs of integrable many-body systems can be derived
by Hamiltonian reduction utilizing the following mechanism.
Suppose that we have a higher dimensional `master phase space' $\cM$ that admits a symmetry group $G$,
and two distinguished independent Abelian Poisson algebras $\fH^1$ and $\fH^2$ formed by $G$-invariant, smooth
functions on $\cM$. Then we can apply Hamiltonian reduction to $\cM$ and obtain
a reduced phase space ${\mathcal M}_{\mathrm{red}}$ equipped with two Abelian Poisson algebras
$\fH_{\mathrm{red}}^1$ and $\fH_{\mathrm{red}}^2$ that descend respectively from
 $\fH^1$ and $\fH^2$.
  We need to construct two distinct models $M$ and $\hat M$ of ${\mathcal M}_{\mathrm{red}}$ yielding $(M,\omega, \fH, \fP)$ and $(\hat M, \hat\omega, \hat\fH, \hat\fP)$
in such a way that the reduction of $\fH^1$ is represented by $\fH$ and $\hat\fP$,
and the reduction of $\fH^2$ is represented by  $\fP$ and $\hat\fH$.
If this is achieved, then we obtain a natural map
$\cR\colon M \to \hat M$ that corresponds to the identity map on  ${\mathcal M}_{\mathrm{red}}$
and relates the Abelian Poisson algebras on $M$ to those on $\hat M$
in the way stated in (\ref{I2}).
A crucial, and very intricate,  requirement
is that the reduction must provide many-body systems: to fulfil this, one can rely only on experience and inspiration.
The heart of the matter is the choice of the correct master system
and its specific reduction.
The examples so far treated by the mechanism just outlined
include group theoretic reinterpretations of dual pairs previously constructed by direct methods
as well as
new dual pairs found by reduction.
At the same time, there still exist such known instances of dualities as well
(notably,  the self-dual hyperbolic RS system \cite{SR88}  and the dual pair involving the
relativistic Toda system \cite{SR90})
that stubbornly resist treatment in the reduction framework.

\begin{figure}[h!]
\centering
\begin{tikzcd}
&{\mathcal M}_0 \arrow{ld}[swap]{\psi}  \arrow{d}{\pi_0}  \arrow{rd} {\hat\psi} \arrow[hook]{r}{\iota_0}
&\cM\\
M\arrow{d}{\lambda} \arrow[bend right]{rr}[swap]{\mathcal R}& {\mathcal M}_{\mathrm{red}} \arrow{r}{\hat\Psi} \arrow{l}[swap]{\Psi} & \hat M\arrow{d}{\hat\lambda}\\
{\mathbb R}^n & & {\mathbb R}^n
\end{tikzcd}
\qquad\qquad
\begin{tikzcd}
&\iota_0^*(\fH^1)\times\iota_0^*(\fH^2)   \\
\fH\times\fP\arrow{ru}{\psi^*} \arrow{r}{\Psi^*} & \fH^1_{\mathrm{red}}\times\fH^2_{\mathrm{red}}\arrow{u}[swap]{\pi_0^*} & \hat \fP\times\hat\fH \arrow{l}[swap]{\hat\Psi^*}\arrow{lu}[swap]{\hat\psi^*}  \arrow[bend left]{ll}{\mathcal R^*}\\
\end{tikzcd}
\caption{ Illustration of how symplectic reduction is used to generate duality. These  diagrams are designed
to help keep track of the notations. Using the embedding $\iota_0\colon \cM_0 \to \cM$
of the `constraint surface' $\cM_0$ into the master phase space $\cM$, the reduced Abelian algebras are defined by
$\fH^i_\red \circ \pi_0 = \fH^i\circ \iota_0$ for $i=1,2$. They turn into the Abelian algebras of the models
$M$ and $\hat M$ according to
$\fH\circ\Psi=\fH^1_{\mathrm{red}}=
\hat\fP\circ\hat\Psi$ and $\fP\circ\Psi=\fH^2_{\mathrm{red}}=\hat\fH\circ\hat\Psi$.
 }
\label{figure X}
\end{figure}

The crucial advantage of the above outlined
approach to action-angle dualities is that, once the correct starting point is found, the
Hamiltonian reduction \emph{automatically} gives rise to complete flows
and symplectomorphisms between
the models of the reduced phase space.
For the realisation  of this advantage, it is indispensable
to provide globally valid descriptions of the
reduced system, which can be a thorny issue.
   The solution of such global issues is at the heart of our current investigation.

The goal of this paper is to present a thorough analysis of a dual  pair of integrable
many-body systems recently derived  in \cite{FG} and \cite{FM} by reduction of the Heisenberg double of
the standard Poisson-Lie group $\SU(2n)$. It is well-known \cite{STS,STSlectures} that the Heisenberg doubles
are Poisson-Lie analogues (and deformations) of corresponding cotangent bundles.
The relevant reduction is a direct Poisson-Lie generalization---making use of Lu's momentum map, \cite{Lu}---of the reduction of the cotangent bundle $T^*\SU(2n)$ used for deriving
the trigonometric $\BC_n$ Sutherland system and its dual in \cite{FG-JMP}.
Correspondingly, the reduction
of the Heisenberg double leads to a deformation of this dual pair.
We shall not only describe the deformed dual pair, but shall also show
how  duality allows  us to extract non-trivial information about the dynamics.
For example, it will allow us to prove that both of the resulting integrable
many-body Hamiltonians are non-degenerate since their flows
densely fill the corresponding Liouville tori.
 Furthermore, it will be shown that all the flows of $\fH$ posses a common fixed point, as do the flows of $\hat\fH$.
 These results will be established by utilizing  the global descriptions of the dual models
  $M$ and $\hat M$ of the reduced phase space.

Our current line of research was initiated in the paper \cite{M}, where the analogous
reduction of the Heisenberg double of $\SU(n,n)$ was considered. The  investigation
in \cite{FG-JMP} was strongly influenced by the work of Pusztai \cite{P}, who studied a dual pair
arising from reduction
of $T^*\SU(n,n)$. The Poisson-Lie counterpart of  the $\SU(n,n)$ dual pair appears more complicated
than what we  report on  here; its exploration is left for the future.

Before outlining the content of the paper, let us recall from \cite{FG,FM}
the local description of our many-body systems in duality, which
arises by restricting attention to dense open submanifolds of the reduced phase space.
These systems have 3 real parameters, $\mu>0$ and $u$ and $v$, whose range will be specified below.
Here, we use hatted letters to describe the model  constructed in \cite{FG}.
 The manifold $\hat M$ contains a dense open proper subset $\hat M^o$
 parametrized by  the Cartesian product
\be
\widehat{\cD}_+ \times \T^n = \{ (\hat\lambda, \exp({\ri \hat\theta}))\},
 \label{I3}\ee
 where $\T^n$ is an $n$-torus and
\be
\widehat{\cD}_+ =\{\hat\lambda\in\R^n\,\mid
 \mathrm{min}(0,v-u)>\hat\lambda_1>\dots >\hat\lambda_n,\,\,\, \hat\lambda_j-\hat\lambda_{j+1}>\mu,\,\, j=1,\dots,n-1\}.
\label{I4}
\ee
The $\hat\lambda_i$ and the angles $\hat\theta_i$ are Darboux coordinates , i.e.,  on $\hat M^o$ we have
 \be
 \hat\omega =\sum_{j=1}^nd\hat\theta_j\wedge d\hat\lambda_j.
\label{I5}\ee
The main Hamiltonian $\hat H$  can be written on $\hat M^o$ as
\be
\hat H(\hat\lambda, \hat\theta)=U(\hat\lambda) - \sum_{j=1}^n\cos(\hat\theta_j) U_1(\hat\lambda_j)^{1/2}
\prod_{\substack{k=1\\(k\neq j)}}^n
\bigg[1-\frac{\sinh^2\mu}{\sinh^2(\hat\lambda_j-\hat\lambda_k)}
\bigg]^{1/2}
\label{I6}\ee
with
\be
\ba
U(\hat\lambda)&=\frac{e^{-2u}+e^{2v}}{2}\sum_{j=1}^n\exp({-2\hat\lambda_j}),\\
U_1(\hat\lambda_j) &= \big[1-(1+e^{2(v-u)})\exp({-2\hat\lambda_j})
+ e^{2(v-u)}\exp({-4\hat\lambda_j})\big].
\ea
\label{I7}\ee
The phase space $M$ of the `dual model' possesses a dense open proper subset $M^o$ parametrized by
\be
\cD_+\times \T^n =\{(\lambda, \exp({\ri \theta}))\}
\label{I8}\ee
with
\be
\cD_+ = \{ \lambda\in \R^n\mid \lambda_1 > \dots > \lambda_n >
\operatorname{max}(\vert v\vert, \vert u \vert),\,\,\, \lambda_{j} - \lambda_{j+1}> \mu,\,\, j=1,\ldots, n-1\}.
 \label{I9}\ee
It carries the Darboux form
\be
 \omega = \sum_{j=1}^nd \theta_j\wedge d\lambda_j.
\label{I10}\ee
In terms of these variables, the main Hamiltonian $H$ reads
 \be
\ba
 H(\lambda,\theta)&=
V(\lambda) +  e^{v-u}\sum_{j=1}^n\frac{\cos\theta_j}{\cosh^2\lambda_j}
\left[1 - \frac{\sinh^2v}{\sinh^2\lambda_j}\right]^{1/2} \left[1 - \frac{\sinh^2u}{\sinh^2\lambda_j} \right]^{1/2}\\
&\qquad\times
\prod_{\substack{k=1\\(k\neq j)}}^n \left[1 - \frac{\sinh^2\mu}{\sinh^2(\lambda_j - \lambda_k)}\right]^{1/2}
\left[1 - \frac{\sinh^2\mu}{\sinh^2(\lambda_j + \lambda_k)}\right]^{1/2}
\ea
\label{I11}\ee
with
\be
V(\lambda) =e^{v-u}\left(\frac{\sinh(v)\sinh(u)}{ \sinh^2\mu}
\prod_{j=1}^n\left[1 - \frac{\sinh^2\mu}{\sinh^2\lambda_j} \right]
-\frac{\cos( v)\cosh(u)}{\sinh^2\mu}
\prod_{j=1}^n\left[1 + \frac{\sinh^2\mu}{\cosh^2\lambda_j} \right]
+ C_0\right)
\label{I12}\ee
where $\displaystyle{ C_0= ne^{u-v} + \frac{\cosh( v -u)}{\sinh^2\mu}}$.
 The constant $C_0$ is included here for later convenience.

The formulae of the main Hamiltonians $\hat H$ (\ref{I6}) and $H$ (\ref{I11}) are invariant with respect to the independent transformations
$\mu \mapsto -\mu$
and $(u,v) \mapsto ( -v, -u)$.
Motivated by this, we assume throughout the paper that $\mu>0$ and at a later stage we shall also assume
that
\be
\vert u \vert > \vert v \vert.
\label{I13}\ee
The exclusion of $\vert u \vert = \vert v \vert$ is  required for our reduction treatment,
while the choice (\ref{I13}) turns out to have technical advantages.
The above specified domains $\widehat\cD_+$ and $\cD_+$  emerge from the reduction,
but they can also be viewed as choices
made to guarantee the strict positivity of all expressions under the square roots
appearing in the Hamiltonians.

A few remarks are now in order. The
main Hamiltonians $\hat H$ and $H$ are reminiscent of many-body Hamiltonians introduced by
van Diejen \cite{vD1}. The relation regarding $\hat H$ was made precise in \cite{FG} and regarding $H$
it will be described in this paper.
The coordinates $\hat\lambda_i$ and $\lambda_i$  serve as position variables
for $\hat H$ and $H$, respectively, and we shall see that they yield globally smooth
(and analytic) functions on the underlying phase space.
Note that
the deformation parameter that brings this dual pair into the one obtained by reduction
of $T^* \SU(2n)$  \cite{FG-JMP}
is here set to unity. The cotangent bundle limits of $\hat H$ and $H$ are discussed in
\cite{FG} and in \cite{FM}.

Now we outline the content of the paper and highlight our main results.
In Section 2.1, we first recall the Heisenberg double $\cM$ equipped with the Abelian
Poisson algebras $\fH^1$ and $\fH^2$,  then set up the pertinent reduction.
In Section 2.2,   we review
 the global model $\hat M$ of the reduced phase space found in \cite{FG}.
 The material in Section 2
 enhances several previous results.
For instance, Lemma 2.1 and the relation (\ref{T44}) of
 $\cH_j^\red$ to Chebyshev polynomials appear here for the first time.
Section 3 contains the logical outline of the construction of
 the global model $M$, which is our primary  task.
 This is summarized by  Figure 2 at the end of Section 3.
 The elaboration of the details required new ideas and a certain amount of labour:
  it occupies Section 4, Section 5 and Section 6.1.
 Our first main result is Theorem 5.6 in Section 5.
 Crucially, this theorem establishes the range of the
 $\lambda$-variables that arises from the reduction.
 Building on  the local results of \cite{FM}, it also yields the Darboux chart (\ref{I10})
 on a dense open submanifold of $\cM_\red$ parametrized by (\ref{I9}).
Our second main  result is given by Theorem 6.5, which
describes the symplectomorphism $\Psi$ between $(M,\omega)$, cast as $\C^n$ with its canonical symplectic structure,  and
$(\cM_\red, \omega_\red)$.
Combining Theorem 6.5 with previous developments, we explain in Section 6.2 that
our reduction engenders a realization of the diagrams of Figure 1.
We  consider this to be our principal achievement.
We also present consequences for the dynamics of the systems in duality
in Section 6.2 and in Section 7.   Section 7 is devoted to further discussion
of the results and open problems.
Finally, two appendices are included. The first one is purely technical, while
in the second we clarify the connection between the Hamiltonian $H$ (\ref{I11}) and van Diejen's
five parametric integrable trigonometric Hamiltonians.

\section{Preparations}
\label{sec:2}

In this section
 we set up the reduction of our interest and review the model $\hat M$ of the
reduced phase space.
All manifolds in this article are viewed as real. Hence the expression ``analytic'' must
always be understood to mean ``real-analytic''.
We shall focus on the $C^\infty$ character of the manifolds and  maps of our concern, but shall
often also indicate their analytic nature by parenthetical remarks.

\subsection{The master system and its reduction}

We shall reduce the master phase space $\cM := \SL(2n,\C)$. Here, $\SL(2n,\C)$ is  viewed as a real Lie group,
and we also need its subgroups
\be
K:= \SU(2n),\quad B:= \SB(2n),
\label{T1}\ee
where the latter is formed by  upper triangular complex matrices with positive entries along the diagonal.
Every element $g\in \cM$ admits the alternative Iwasawa decompositions
\be
g = k b = b_L k_R,
\qquad k, k_R \in K, \quad b, b_L \in B.
\label{T2}\ee
By using these, $\cM$ is equipped with the Alekseev-Malkin \cite{AM} symplectic form
\be
\omega_\cM=\frac{1}{2}\Im\tr(db_Lb_L^{-1}\wedge  d k k^{-1})+
\frac{1}{2}\Im\tr( b^{-1}db  \wedge k_R^{-1}  d k_R).
\label{T3}\ee
To display the corresponding Poisson bracket, for any
$\cF\in C^\infty(\cM, \R)$ we introduce the $\sl(2n,\C)$-valued
left- and right-derivatives $\nabla \cF$ and $\nabla' \cF$ by
\be
\ds \cF(e^{sX}ge^{sY})=\Im\tr\big(X \nabla \cF(g) +Y \nabla' \cF(g)\big),
\quad\forall X,Y\in\sl(2n,\C).
\label{T4}\ee
We prepare the linear operator
\be
R = \frac{1}{2} (\pi_\cK - \pi_{\cB})
\label{T5}\ee
on $\sl(2n,\C)$, utilizing the projectors associated with the real vector space decomposition
\be
\sl(2n,\C) = \cK + \cB,
\label{T6}\ee
where $\cK$ and $\cB$ are the Lie algebras of $K$ and $B$, respectively.
The Poisson bracket   reads
\be
\{ \cF, \cH\} = \Im\tr\left( \nabla \cF R(\nabla \cH) + \nabla'\cF R(\nabla' \cH) \right),
\qquad
\cF, \cH\in C^\infty(\cM,\R).
\label{T7}\ee
The structure described above is known \cite{STS,STSlectures} as the Heisenberg double
of the standard Poisson-Lie group $\SU(2n)$.

The Abelian Poisson algebra $\fH^2$ is defined as follows.
Let $\cP$ denote the space of positive definite Hermitian matrices of size $2n$ and determinant $1$.
Consider the ring $C^\infty(\cP)^K$ of smooth real function on $\cP$ that are invariant with respect
to the natural action of $K$ on $\cP$ given by conjugation of a Hermitian matrix by a unitary one.
We set
\be
\fH^2 = \{ \hat\cH \in C^\infty(\cM)\mid \hat \cH(g) = \hat h( b b^\dagger) \,\,\hbox{with}\,\,
\hat h \in C^\infty(\cP)^K\},
\label{fH2}\ee
i.e., $\fH^2$ is the pull-back of $C^\infty(\cP)^K$ by the map $ \cM \ni g\mapsto bb^\dagger \in \cP$.
A generating set $\hat \cH_j$ for $\fH_\cM^2$ is provided by the functions $\hat \cH_j$ having the form
\be
\hat \cH_j(g) = \hat h_j(bb^\dagger) \quad\hbox{with}\quad
\hat h_j(bb^\dagger):=  \frac{1}{2}\tr\!\left((bb^\dagger)^j \right)\quad\hbox{ for}\quad  j=1,\dots, 2n-1.
\label{T9}\ee
The Hamiltonian vector field and the corresponding (complete) flow can be  written down explicitly for any
$\hat \cH\in \fH^2$.
 After our reduction the $n$ Hamiltonians descending from the functions  $\hat\cH_1, \hat\cH_2,\dots,\hat\cH_n$
 remain independent,  and
the many-body Hamiltonian displayed in
(\ref{I6}) results from $\hat \cH_1$.

To present the other Abelian Poisson algebra of interest, $\fH^1$,
we define the matrix
\be
I:= \diag(\1_n, - \1_n),
\label{T10}\ee
where $\1_n$ is the $n\times n$ unit matrix, and introduce the subgroup
\be
K_+ := \{ k \in K \mid k^\dagger I k = I\}.
\label{T11}\ee
Let $C^\infty(K)^{K_+ \times K_+}$ denote those functions on $K$ that are invariant
 with respect to both left- and right-multiplications by elements of $K_+$.
Then,   referring to the Iwasawa decomposition (\ref{T2}),
we define
\be
\fH^1 = \{ \cH \in C^\infty(\cM)\mid  \cH(g) =  h(k) \,\,\hbox{with}\,\,
h \in C^\infty(K)^{K_+\times K_+}\}.
\label{fH1}\ee
A generating set is furnished  by the functions $\cH_j$ given by
\be
\cH_j(g) = h_j(k) \quad \hbox{with}\quad h_j(k):=\frac{1}{2}
 \tr\!\left( (k^\dagger I k I)^j\right) \quad \hbox{for}\quad j=1,\dots, n.
\label{T13}\ee

We recall that $C^\infty(K)$  carries a  natural Poisson bracket associated with (\ref{T7}), for which
the map $g\mapsto k$ by (\ref{T2}) is a Poisson map. Explicitly,
\be
\{f,h\}_K(k) =\Im\tr \left(D f(k) k (D' h(k)) k^{-1}\right),\quad \forall k\in K,\, f,h\in C^\infty(K).
\label{T14}\ee
Here the $\cB$-valued left- and right-derivatives, $Df$ and $D'f$, of any $f\in C^\infty(K)$ are defined analogously to (\ref{T4}).
It is well-known that $K$ is a Poisson-Lie group  and
 $K_+ < K$ is a Poisson-Lie subgroup of $K$  with respect to this Poisson structure.
The following lemma implies  that $\fH^1$ is an Abelian Poisson algebra.

\medskip\noindent
{\bf Lemma 2.1.} \emph{The invariant functions $C^\infty(K)^{K_+ \times K_+}$
Poisson commute with respect to
$\{\ ,\ \}_K$.}

\medskip\noindent{\bf Proof.}
Let us start by noting that every $k\in K$ may be written in
the form
\be
k = \kappa_1 \Delta \kappa_2,\qquad \hbox{for}\ \kappa_1, \kappa_2\in K_+, \ \hbox{and}\
\Delta=\bpm\Gamma& \ri\Sigma\\ \ri\Sigma&\Gamma\epm,
\label{T15}\ee
where
\be
\Gamma=\diag(\cos q_1,\dots,\cos q_n),\qquad  \Sigma=\diag(\sin q_1,\dots,\sin q_n)
\label{T16}\ee
with
\be
\frac{\pi}{2}\geq q_1 \geq q_2\geq \dots \geq q_n\geq 0.
\ee
If $h_1$ and $h_2$ are two $(K_+\times K_+)$-invariant smooth functions on $K$, then their Poisson bracket is
also $(K_+\times K_+)$-invariant.
 Therefore it is enough to show that $\{h_1,h_2\}$ vanishes at any point of the form $\Delta$ given in (\ref{T15}).

The $(K_+\times K_+)$-invariance of $h\in C^\infty(K)$  means that the $\cB$-valued
left- and right-derivatives
$D h, D' h$ have the form
\be
D h=\bpm
0&A\\ 0&0
\epm,
\qquad
D' h=\bpm
0&\tilde A\\ 0&0
\epm,
\label{T18}\ee
where we use the obvious $2\times 2$ block-structure defined by $I$ (\ref{T10}).
 On account of the identity
\be
D h(k)=\pi_\cB(k(D'h(k))k^\dag) = k (D' h(k)) k^\dag - \pi_\cK(k(D'h(k))k^\dag),
\label{T19}\ee
 we must also  have
\be
 D'h(k) - k^\dag ( D h(k)) k \in\cK.
 \label{T20}\ee
Applying this at $k=\Delta$, we obtain
\be
\bpm
-\ri\Gamma A \Sigma & \tilde A  - \Gamma  A \Gamma\\
-\Sigma  A  \Sigma & \ri\Sigma  A \Gamma
\epm
\in\cK,
\label{T21}\ee
where the dependence of $A$ and $\tilde A$ on $\Delta$ is suppressed.
This gives us the conditions (the first two from skew symmetry of the diagonal blocks,
and the third---after a calculation---from comparison of the off-diagonal blocks)
\be
\ba
&(i)\qquad\,\, \Sigma A \Gamma = \Gamma  A^\dag\Sigma,\\
&(ii)\qquad \Gamma A \Sigma  = \Sigma  A^\dag \Gamma ,\\
&(iii)\qquad\,\, \Gamma\tilde A  =  A \Gamma.
\ea
\label{T22}\ee

Let $h_1$ and $h_2$ be two  $(K_+\times K_+)$-invariant functions, and use
$A_1$, $\tilde A_1$  and $A_2$, $\tilde A_2$ as in (\ref{T18})
for their derivatives.
By substitution into the Poisson bracket (\ref{T14}) on $K$ we get
\be
\{h_1,h_2\}_K(\Delta) =
\Im\,\tr\left(\Sigma  A_1 \Sigma \tilde A_2\right),
\label{T23}\ee
which
then produces
$\{h_1,h_2\}_K(\Delta)
= \Im\,\tr \left(A_1\Sigma\Gamma^{-1}  A_2 \Gamma\Sigma\right)$
using $(iii)$ of (\ref{T22}).  Utilizing alternatively $(ii)$ and $(i)$ then gives
\be
\{h_1,h_2\}_K(\Delta)=
\Im\,\tr\left( \Sigma^2 A_1A_2^\dag\right)\quad\hbox{and}\quad
\{h_1,h_2\}_K(\Delta)=
\Im\,\tr\left(\Sigma^2 A_1^\dag A_2\right).
\label{T24}\ee
The combination of $(i)$ and $(ii)$ yields $\Gamma^2  A \Sigma^2 = \Sigma^2 A \Gamma^2$,
and thence $[\Sigma^2,A]=0$.
 Applying this to the two expressions in (\ref{T24}) and then adding them, we have
\be
\ba
2\{h_1,h_2\}_K(\Delta)&=\Im\,\tr\left( A_1\Sigma^2A_2^\dag\right) + \Im\,\tr\left(\Sigma^2 A_1^\dag A_2\right)\\
&=
-\Im\,\tr\left(A_2\Sigma^2A_1^\dag\right) + \Im\,\tr\left(\Sigma^2 A_1^\dag A_2\right)
=
\Im\,\tr\left(A_1^\dag[A_2,\Sigma^2]\right)=0,
\ea
\label{T25}\ee
which completes the proof.
\hfill $\square$
\medskip

The Hamiltonian vector fields corresponding to the collective Hamiltonians  $\cH\in \fH^1$ (\ref{fH1})
are all complete.
Actually the completeness is valid
for any $\cH\in C^\infty(\cM)$  given by $\cH(g) = h(k)$ using the Iwasawa decomposition $g=kb$ (\ref{T2})
and any $h\in C^\infty(K)$. In this case the derivatives of $\cH$ are related to the derivatives of $h$
according to
\be
\nabla' \cH(g) = b^{-1} (D' h(k)) b \, \in\cB,\qquad
\nabla \cH(g) = k (D' h(k)) k^{-1}.
\label{T26}\ee
This implies that the integral curves $g(t) = k(t) b(t)$ of the Hamiltonian vector field of $H$ on $\cM$
are determined by the `decoupled' differential equations
\be
\dot{k}= \pi_\cK( k (D' h(k)) k^{-1}) k \quad\hbox{and}\quad \dot b = - (D'h(k)) b.
\label{T27}\ee
The vector field on $K$ occurring in the first equation is complete due to compactness of $K$.
After substituting  a solution $k(t)$ into the second equation, $b(t)$ can be found (in principle) by
performing a finite number of integrations: this is because of the triangular structure of the group $B$.

Now, with the master phase space $\cM$ and its two distinguished Abelian Poisson
 algebras $\fH^1$ and $\fH^2$ at our
disposal, we summarize the reduction procedure that concerns us.
The basic steps of defining a reduction are the specifying of the symmetry
and of the constraints to be used.
As our symmetry group, we take the direct product $K_+ \times K_+$ and let it act
on the phase space by the map
\be
\Phi\colon K_+ \times K_+ \times \cM \to \cM,
\qquad
(\eta_L, \eta_R, g) \mapsto \eta_L g \eta_R^{-1}.
\label{T28}\ee
This is a Poisson action if $K_+$ is endowed with its
natural multiplicative Poisson structure inherited from (\ref{T14}) \cite{STS,STSlectures}.
The momentum map generating this action
 sends $g$ to the pair of matrices given by the block-diagonal parts of $b_L$ and $b$ (\ref{T2}).
The constraints restrict the value of the momentum map to a suitable constant.
To define the constraints, we fix a  positive number $\mu$ and a vector $\hat v \in \C^n$,
and  let $\sigma$ denote the unique upper triangular matrix with positive diagonal entries
that verifies
\be
\sigma \sigma^\dagger = \alpha^2 \1_n +
\hat v \hat v^\dagger, \quad \hat v^\dagger \hat v = \alpha^2 (\alpha^{-2n} - 1), \quad \alpha:= e^{-\mu}.
\label{T29}\ee
Then we impose the  `left-handed' momentum map constraint forcing $b_L$ to have the form
\be
b_L =
\begin{bmatrix} y^{-1}\sigma  & \chi_L \\
0  &  y \1_n  \end{bmatrix},
\qquad y = e^{-u} ,
\label{T30}\ee
 and also impose the  `right-handed' momentum map constraint by requiring that $b$ reads
\be
b =
\begin{bmatrix} x \1_n  &  \chi \\
0  &  x^{-1} \1_n  \end{bmatrix},\qquad
x = e^{-v}
\label{T31}\ee
with real parameters $u$ and $v$ subject to
$\vert u \vert \neq \vert v \vert$.
We use a $2\times 2$ block-matrix notation corresponding to $I$ (\ref{T10}), and thus $\chi_L$, $\chi$ are
$n\times n$ complex matrices.
The submanifold $\cM_0$ of $\cM$ defined by these momentum constraints,
\be
\cM_0=\{g\in\cM\ |\ b_L(g)\ \hbox{and}\ b(g)\ \hbox{determined by
(\ref{T30}) and (\ref{T31})}\},
\label{defM0}\ee
 is stable under the action of the
`gauge group' $K_+(\sigma) \times K_+$, where
\be
K_+(\sigma):= \{ \eta_L\in K_+ \mid  \eta_L \diag(\sigma \sigma^\dagger, \1_n) \eta_L^{-1}
=  \diag(\sigma \sigma^\dagger, \1_n) \}.
\label{T32}\ee
According to general principles, the reduced phase space $\cM_\red$ is the quotient
\be
\cM_\red = \cM_0/ (K_+(\sigma) \times K_+).
\label{T33}\ee
It was shown in \cite{FG} that the `effective gauge group'
\be
(K_+(\sigma) \times K_+)/ \Z_{2n}^\diag
\label{T34}\ee
acts \emph{freely} on $\cM_0$, where
$\Z_{2n}^\diag$ is the subgroup that acts trivially
\be
\Z_{2n}^\diag =\{ (\zeta \1_{2n}, \zeta \1_{2n}) \in K_+(\sigma)\times K_+ \mid \zeta \in \Z_{2n}\}.
\label{T35}\ee
In other words, $\pi_0\colon \cM_0 \to \cM_\red$ is a principal fibre bundle with structure group (\ref{T34}).
It follows that $\cM_\red$  is a smooth (and analytic) symplectic manifold,
and we let $\omega_\red$ denote its symplectic form that descends from $\omega_\cM$.
It is readily seen that all elements of
$\fH^1$ and $\fH^2$
 are invariant with respect to the group action
$\Phi$ (\ref{T28}) on $\cM$,
and thus they give rise to two Abelian Poisson algebras
$\fH_\red^1$ and $\fH_\red^2$
 on the symplectic manifold $(\cM_\red, \omega_\red)$.
 Referring to equations (\ref{T9}), (\ref{T13}) and using the embedding $\iota_0\colon\cM_0\to \cM$ as in Figure 1,
 the defining relations of the reduced Hamiltonians of our principal interest are
\be
\cH_j^\red\circ  \pi_0 = \cH_j \circ \iota_0, \quad \hat \cH_j^\red\circ \pi_0 = \hat \cH_j \circ \iota_0,
\label{redhams}\ee
and of course $ \pi_0^*(\omega_\red)= \iota_0^\star (\omega_\cM)$.
In the spirit of the general scheme outlined in the Introduction,
our task now is to construct a suitable pair of models of $\cM_\red$.
 One model was already found before, and next we briefly recall it.

\subsection{The model $\hat M$ of $\cM_\red$ and its consequences}

 The construction presented in this subsection is extracted from \cite{FG}, where  details can be found.

As the first main step,
a parametrization of  a dense open submanifold of the reduced phase space
by the domain $\widehat\cD_+ \times \T^n$ (\ref{I4}) was constructed, where the variables $\hat\lambda_i$ are related
to the invariant  $\Delta$ (\ref{T15}) formed from  $k$ in
$g=kb\in \cM_0$  by setting
\be
\sin q_i = \exp({\hat\lambda_i}),
\label{T36}\ee
using that $q_n>0$ for $g\in \cM_0$.
It proves useful to combine the
$\hat\lambda_i\in{\mathbb R}_{<0}$ and their
 canonical conjugates $\hat\theta_i\in{\mathbb R}/2\pi{\mathbb Z}$
 into
complex variables  by defining
\be
\cZ_j(\hat\lambda,\exp({\ri\hat\theta}))
=(\hat\lambda_j-\hat\lambda_{j+1}-\mu)^{\tfrac{1}{2}}\prod_{k=j+1}^n \exp({\ri\hat\theta_k}),
\quad j=1,\dots,n-1,
\label{T37}\ee
and
\be
\cZ_n(\hat\lambda,\exp({\ri\hat\theta}))=(s-{\hat\lambda_1})^{\tfrac{1}{2}}\prod_{k=1}^n\exp({\ri\hat\theta_k})\quad\hbox{with}
\quad s=\operatorname{min}(0, v-u).
\label{T38}\ee
The variable $\cZ$ is naturally extended ro run over the whole of $\C^n$, equipped with
 the symplectic form
\be
 \omega_\can = \ri\sum_{j=1}^{n}d \cZ_j\wedge d  \cZ_j^*.
\label{T39}\ee
The domain $\widehat\cD_+ \times \T^n$ with (\ref{I5}) is
symplectomorphic to the dense open subset $(\C^*)^n$ of $\C^n$.
The main result  of \cite{FG} says that
\be
(\hat M,  \hat \omega) \equiv (\C^n,  \omega_\can)
\label{T40}\ee
is a model of the \emph{full reduced phase space $(\cM_\red, \omega_\red)$ (\ref{T33})}.
In fact,  one can construct a symplectomorphism
\be
\hat \Psi\colon \cM_\red \to \hat M, \qquad \hat \Psi^*(\hat \omega) = \omega_\red.
\label{T41}\ee

The $n$-tuples $(\hat\lambda_1,\dots, \hat\lambda_n)$ and $(\vert \cZ_1\vert^2,\dots, \vert \cZ_n\vert^2)$
yield  analytic maps from  $\hat M$ to  $\R^n$, which
are related by an affine $\GL(n,\Z)$ transformation. Explicitly, we have
\be
\hat\lambda_j(\cZ)=s-(j-1)\mu - |\cZ_n|^2 -\sum_{l=1}^{j-1} \vert \cZ_l \vert^2,
\qquad j=1,\dots, n.
\label{T42}\ee
The functions $\vert \cZ_j \vert^2$ generate  the obvious Hamiltonian action
of the torus $\T^n$ on $\hat M=\C^n$.  Namely, the flows of $\vert \cZ_1\vert^2,\dots, \vert \cZ_n\vert^2$
with time parameters
$t_1,\dots, t_n$ act by the map
\be
(\cZ_1,\dots, \cZ_n) \mapsto (\cZ_1e^{\ri t_1},\dots, \cZ_n e^{\ri t_n}).
\label{T43}\ee
The origin $\cZ_1 = \dots = \cZ_n =0$ is the unique fixed point of this action.
Applying  (\ref{T15}) and (\ref{T36}), the reduced Hamiltonians
$\cH_j^\red\in C^\infty(\cM_\red)$ that descend from the functions
 $\cH_j$ (\ref{T13}) are found to
take the following form in terms of the model $\hat M$:
\be
\cH_j^\red\circ \hat \Psi^{-1} =  \sum_{i=1}^n P_j(\exp({ \hat\lambda_i})),
\label{T44}\ee
where $P_j$ is the polynomial determined by the relations
\be
\cos(2j q_a) = P_j( \exp({ \hat\lambda_a})),
\quad
\exp({\hat\lambda_a})=\sin q_a  \quad\hbox{for}\quad 0< q_a \leq \frac{\pi}{2}.
\label{T45}\ee
 That is,
\be\ba
P_j(\exp({\hat\lambda_a})) &= T_j\bigl(\cos(2 q_a)\bigr) = T_j\bigl(1-2\sin^2(q_a)\bigr) =
  (-1)^jT_j\bigl(2\sin^2(q_a)-1\bigr) \\
&= (-1)^jT_{2j}\bigl(\sin(q_a)\bigr)=(-1)^jT_{2j}\bigl(\exp({\hat\lambda_a})\bigr) ,
\ea
\label{T46}\ee
where  $\{T_m(x) \mid m=0,1,2,\dots\}$ are Chebyshev polynomials of the first kind, characterized by
$
T_m(\cos\varphi)=\cos(m\varphi)$.

Altogether, we see that the $\hat\lambda_j$, or  equivalently the $\vert \cZ_j\vert^2$,
are action variables
for the Liouville integrable system defined by the reduced Hamiltonians  $\cH_1^\red,\dots, \cH_n^\red$.
The subset of $\hat M$ on which $\prod_{j=1}^n \cZ_j=0$ is mapped by (\ref{T42})
onto the boundary of the closure of $\widehat\cD_+$, with
$\cZ=0$ corresponding to the vertex
\be
\hat\lambda_j= s - (j-1)\mu, \qquad j=1,\dots, n, \quad s=\min(0,v-u).
\label{T47}\ee
The point $\cZ=0$ is a common equilibrium for the Hamiltonians $\cH_j^\red\circ \hat \Psi^{-1}$.
Moreover, $\cH_1^\red\circ \hat \Psi^{-1}$ reaches its global minimum on $\hat M$ at $\cZ=0$.
This  follows from the fact that $\cos(2 q_a)$ is monotonically
decreasing for $ 0 < q_a \leq \frac{\pi}{2}$ and the joint maxima of the $q_a$ for $a=1,\dots, n$ is reached
at the vertex (\ref{T47}) corresponding to $\cZ=0$.

On the dense open  subset parametrized by  $\widehat\cD_+\times \T^n$, the flow generated by
$\cH_j^\red\circ \hat \Psi^{-1}$ (\ref{T44})
 is linear
\be
\hat\lambda_a(t_j) = \hat\lambda_a(0),\quad
\hat\theta_a(t_j) = \hat\theta_a(0) + t_j \hat \Omega_{j,a}(\hat\lambda_a),\qquad a=1,\dots, n,
\label{T49}\ee
with the frequencies
\be
\hat \Omega_{j,a}(\hat\lambda_a) = \frac{\partial P_j(\exp({ \hat\lambda_a}))}{\partial \hat\lambda_a}
= 2(-1)^j j\, \exp({\hat\lambda_a})U_{2j-1}\bigl(\exp({\hat\lambda_a})\bigr),
\label{T50}\ee
 where $\{U_m(x)\mid m=0,1,2,\dots\}$ are Chebyshev polynomials of the second kind, characterized by
$
U_{m}(\cos\varphi)={\sin((m+1)\varphi)}/{\sin(\varphi)}
$.
It is obvious that for generic $\hat\lambda$ and any fixed $j$ the frequencies
\be
\hat \Omega_{j,1}(\hat\lambda_1),\dots, \hat \Omega_{j,n}(\hat\lambda_n)
\label{T51}\ee
are independent over the field of rational numbers, and therefore
the flow of $\cH_j^\red\circ\hat \Psi^{-1}$ densely fills the generic Liouville tori.
This implies that every element in the commutant of
$\cH_j^\red\circ \hat \Psi^{-1}$ in $C^\infty(\hat M)$  is a
function of
the action variables $\hat\lambda_1,\dots, \hat\lambda_n$. In other words, each
$\cH_j^\red\circ \hat \Psi^{-1}$ is a non-degenerate
completely integrable Hamiltonian.

On the full phase space $\hat M$, the flow generated by the function
$\cH_j^\red\circ \hat \Psi^{-1}$ has the following form:
\bea
&&\cZ_k(t_j) = \cZ_k(0)
\exp\left(\ri t_j \left(\hat \Omega_{j,k+1}(\hat\lambda_{k+1}) +\dots + \hat\Omega_{j,n}(\hat\lambda_n)\right)\right),
\quad k=1,\dots, n-1,\nonumber\\
&&\cZ_n(t_j) = \cZ_n(0)
\exp\left(\ri t_j\left(\hat \Omega_{j,1}(\hat\lambda_1) +\dots + \hat\Omega_{j,n}(\hat\lambda_n)\right)\right),
\label{T52}\eea
where here $\hat\lambda$ is evaluated on the initial value $\cZ(0)$.

  As for the
 reduced Hamiltonians $\hat H_j := \hat \cH_j^\red  \circ \hat\Psi^{-1}$
descending from $\hat \cH_j$ (\ref{T9});
$\hat H \equiv \hat H_1$  takes
the Ruijsenaars--Schneider--van Diejen (RSvD) type many-body form (\ref{I6})
in terms of the variables $(\hat\lambda, \hat\theta)$. This Hamiltonian
as well as all members of its commuting family yield
analytic functions on the full reduced phase space modelled by $\hat M$.
Explicit formulae can be obtained following the lines of  \cite{FG}.
By using its analyticity and the asymptotic behavior where the particles are far apart,
it can be shown  that the determinant
$\det\big[d_{\hat \theta}\hat H_1,d_{\hat \theta}\hat H_2,\dots,d_{\hat \theta}\hat H_n\big]$
is  non-zero on a dense open subset of $\widehat\cD_+ \times \T^n$.
This not only implies the Liouville integrability of the Hamiltonians $\hat H_j$,
but it
shows also that the $2n$ functions
 $\hat\lambda_j\in\hat \fP$ and  $\hat H_j\in \hat \fH$, for  $j=1,\dots, n$,  are functionally independent.
In particular,  the Hamiltonian vector fields of the
elements of $\hat \fP$ and $\hat \fH$ together span the tangent space $T_m \hat M$
at generic points $m\in \hat M$.
As a consequence of the formula  (\ref{T44}), $\{\hat \lambda_j\}_{j=1}^n$
and $\{\cH_j^\red\circ \hat \Psi^{-1}\}_{j=1}^n$ represent
alternative generating sets for the algebra $\hat \fP$ of the global position variables.

\medskip\noindent
{\bf Remark 2.1.}
In \cite{FG} the model $\hat M$ was obtained under the assumptions that $v>u$ and $\vert u \vert \neq \vert v \vert$,
but now we find that essentially nothing
changes if only $\vert u\vert \neq \vert v\vert$ is assumed.
The condition $\hat\lambda_1 \leq s=\mathrm{min}(0,v-u)$ arises from the requirement that all entries
of the $n\times n$ diagonal matrix given by
$K_{22} K_{22}^\dagger = e^{-2u}\1_n - (\sin q)^2 e^{-2v}$
in equation (3.8)
of \cite{FG} must be non-negative.
Another difference is that \cite{FG}
 defined $z_1,\dots, z_{n-1}$
in the same way as (\ref{T37}), but introduced a variable $z_n$
instead of $\cZ_n$  (\ref{T38}) by
\be
z_n(\hat\lambda, \exp({\ri \hat\theta})) = (e^s-e^{\hat\lambda_1})^{\tfrac{1}{2}}\prod_{k=1}^n \exp({\ri\hat\theta_k}),
\label{T53}\ee
which varies in the open disc $\D_r$ of radius $r=e^{s/2}$, and
 is related to $\cZ_n\in \C$ by an analytic diffeomorphism.

\section{Constructing the model $M$ of $\cM_\red$: general outline}
\label{sec:3}

The model $\hat M$ of $\cM_\red$ was obtained
by explicitly constructing a global cross-section
of the gauge orbits in $\cM_0$.  The construction of the new model $M$ that
 we achieve in this paper
is somewhat more complicated. We here collect the main concepts
that will appear in the construction, hoping that this will enhance readability.
The reader is recommended to keep an eye on Figure 2, placed at the end of the section.

We shall describe the quotient $\cM_\red$ (\ref{T33}) by exhibiting a new set of unique representatives
for each orbit of the `gauge group' $K_+(\sigma) \times K_+$ acting on $\cM_0$.
We now display $\cM_\red$ as
\be
\cM_\red = K_+(\sigma) \backslash \cM_0/ K_+ \, ,
\label{P1}\ee
emphasising that $(\eta_L, \eta_R) \in K_+(\sigma) \times K_+$ acts by left- and
by right-multiplication, respectively.
We shall arrange taking the quotient into convenient consecutive steps, using in addition to the obvious direct product
structure of the gauge group also the fact that $K_+(\sigma)$ itself can be decomposed as
the direct product
\be
K_+(\sigma) = K_+(\hat w) \times \T_1,
\label{P2}\ee
where
\be
\T_1 = \{\hat \gamma:= \diag(\gamma \1_n, \gamma^{-1} \1_n)\mid \gamma\in \U(1)\},
\label{P4}\ee
\be
K_+(\hat w)=\{ \kappa \in K_+\mid \kappa \hat w = \hat w \} \quad\hbox{with}\quad  \hat
w= (\hat v,0)^T\in \C^{2n},
\label{P3}\ee
 and $\hat v\in \C^n$ is the fixed vector in (\ref{T29}).
It is easy to check that every element of $K_+(\sigma)$ can be written as a product of these
two disjunct, mutually commuting subgroups.

As in \cite{FM}, we call $b\in B$  `quasi-diagonal' if it has the
form
\be
b =
\begin{bmatrix} e^{-v} \1_n  & \beta \\
0  &  e^v \1_n  \end{bmatrix}
\quad\hbox{with}\quad
\beta= \diag(\beta_1, \ldots, \beta_n), \quad \beta_1  \geq \beta_2\geq \dots \geq \beta_n \geq 0,
\label{P5}\ee
and define the subset $\cM_1$ of the `constraint surface' $\cM_0\subset \cM$ by
\be
\cM_1:= \{ g = k b\in \cM_0\mid \hbox{$b$ is quasi-diagonal}\}.
\label{P6}\ee
The `left-handed' gauge transformations by $K_+(\sigma)$ map $\cM_1$ to itself and by using this
we introduce the quotient
\be
\cN:= K_+(\hat w)\backslash \cM_1.
\label{P7}\ee
It will be useful to identify $\cN$ with the image of the map
\be
\cM_1 \ni g= k b \mapsto (w(k), L(k), \beta)
\quad \hbox{with}\quad w(k):= k^{-1} \hat w,\,\, L(k):= k^{-1} I k I.
\label{P8}\ee
Such an identification is possible since
 $(w(k),L(k))=(w(k'),L(k'))$ for $k, k'\in K$  if and only if $k'k^{-1}\in K_+(\hat w)$.
 Directly from the definitions, we have
$K_+(\sigma)\backslash\cM_1 =\T_1\backslash\cN$, where $\hat \gamma\in \T_1$ (\ref{P4}) acts according to
$w(\hat \gamma k)=\gamma^{-1} w(k)$, because
of the form of $\hat w$ in (\ref{P3}), while $L(k)$
 and $\beta$ are unchanged.

The gauge transformation (\ref{T28}) by  $(\eta_L, \eta_R)\in K_+(\sigma)\times K_+$ acts on the $k$ and $b$ components
of $g=kb\in \cM_0$ by
\be
(k,b) \mapsto (\eta_L k \eta_R^{-1}, \eta_R b \eta_R^{-1}),
\label{P9}\ee
and thus operate on the constituent $\chi$ (\ref{T31}) of $b$  according to
\be
\chi \mapsto \eta_R(1) \chi \eta_R(2)^{-1},
\label{P10}\ee
where we employ the block-matrix notation
\be
\eta_R=
\begin{bmatrix} \eta_R(1)  & 0 \\
0  &  \eta_R(2)  \end{bmatrix} ,
\quad
\eta_R(1), \eta_R(2)\in \U(n), \quad \det(\eta_R(1) \eta_R(2)) =1.
\label{P11}\ee
Recalling the singular value decomposition of $n\times n$ matrices,
we observe from (\ref{P10}) that every element $g\in \cM_0$ can be gauge transformed into $\cM_1$,
and the components $\beta_i$ of the resulting element of $\cM_1$ are uniquely determined by $g$.
To proceed further, we restrict ourselves to the `regular part'   defined
by the strict inequalities
\be
\beta_1 > \beta_2 > \dots >\beta_n >0.
\label{P12}\ee
We call such $\beta$ and the corresponding quasi-diagonal $b$ \emph{regular}, and apply
 the notations  $\cM_1^\reg$, $\cM_0^\reg$, $\cN^\reg$ and $\cM_\red^\reg$ for the corresponding subsets.
Specifically, $\cM_0^\reg$ consists of the elements of $\cM_0$ that can be
gauge transformed into $\cM_1^\reg$, $\cN^\reg = K_+(\hat w)\backslash \cM_1^\reg$ and
$\cM_\red^\reg = K_+(\sigma)\backslash\cM_0^\reg/K_+$.
\emph{ Later it will emerge that in fact $\cM_0 = \cM_0^\reg$.}

If $\beta$ is regular, then the corresponding $b$  in (\ref{P5}) is fixed by the following
Abelian subgroup, $\T_{n-1}$, of $K_+$:
\be
\T_{n-1}:=\{ \delta= \diag(\delta_1,\dots, \delta_n, \delta_1,\dots, \delta_n)\mid
\delta_i\in \U(1),\quad \prod_{i=1}^n \delta_i^2=1\}.
\label{P13}\ee
We shall also use the subgroup of $\T_1 \times  \T_{n-1}$  given by
\be
\tilde \Z_{2n}^\diag =
\{ (\hat \zeta , \zeta \1_{2n})\mid \zeta\in \Z_{2n}\},
\label{P14}\ee
where $\Z_{2n}$ denotes the $(2n)^{\mathrm{th}}$ roots of unity and we  employ the notation (\ref{P4}).
Defining
\be
\T_n := \{ \tau=\diag(\tau_1,\dots, \tau_n, \tau_1,\dots, \tau_n)\mid \tau_i\in \U(1)\},
\label{P15}\ee
we have the isomorphism
\be
 \T_{n} \simeq (\T_1 \times \T_{n-1})/\tilde \Z_{2n}^\diag ,
\label{P16}\ee
which is provided by the  map
\be
\tau_i = \gamma^{-1} \delta_i
\label{P17}\ee
using  the above parametrizations of the elements of $\T_1$ (\ref{P4}), $\T_{n-1}$ (\ref{P13})
and $\T_n$ (\ref{P15}).

After these preparations, we come to the main points.
First, we let $\delta\in \T_{n-1}$ act on $\cM_1^\reg \times K_+$ by
\be
\delta\colon   (g, \eta) \mapsto (g \delta^{-1}, \delta \eta)
\label{P18}\ee
and also let $\eta_R\in K_+$ act by
\be
\eta_R \colon ( g, \eta) \mapsto (g, \eta \eta_R^{-1}).
\label{P19}\ee
Then we introduce the identification
\be
\cM_0^\reg \longleftrightarrow (\cM_1^\reg \times K_+)/\T_{n-1}
\label{P20}\ee
by means of the map
\be
(\cM_1^\reg \times K_+)\ni  (g, \eta) \mapsto g\eta\in \cM_0^\reg,
\label{P21}\ee
which is invariant  with respect to the action (\ref{P18}) of $\T_{n-1}$.
Since the actions of $\T_{n-1}$ and $K_+$ on $\cM_1^\reg \times K_+$ commute, we have
\be
\cM_0^\reg/K_+ = ((\cM_1^\reg \times K_+)/ \T_{n-1})/ K_+  =((\cM_1^\reg \times K_+)/K_+)/ \T_{n-1}= \cM_1^\reg/\T_{n-1},
\label{P22}\ee
where on the right-end we refer to the action of $\T_{n-1}$ on $\cM_1^\reg$ given by
 $\cM_1^\reg \ni k b \mapsto k b\delta^{-1}= k\delta^{-1} b$.
We continue by applying the decomposition (\ref{P2}) of  $K_+(\sigma)$ to deduce the identification
\be
\cM_\red^\reg = ( K_+(\hat w)\times\T_1 )\backslash \cM_1^\reg / \T_{n-1} = \T_1 \backslash \cN^\reg /\T_{n-1},
\label{P23}\ee
where we have taken into account that  $\cN^\reg = K_+(\hat w)\backslash \cM_1^\reg$ (see (\ref{P7})).
The action of $\T_1 \times \T_{n-1}$ on $\cN^\reg$ factors through the homomorphism  (\ref{P17}).
The induced action of $\T_n$ (\ref{P15}) on $\cN^\reg$ is given,
in terms of the triples $(w, L, \beta)$ in (\ref{P8}) representing the elements of $\cN^\reg$, by the formula
\be
 (w,L,\beta) \mapsto (\tau w, \tau L \tau^{-1},\beta),\qquad \forall \tau\in \T_n.
\label{P26}\ee
One sees this from the definitions in (\ref{P8}) and in (\ref{P17}) using that
$(\hat \gamma, \delta) \in \T_1 \times \T_{n-1}$ sends $g = k b\in \cM _1^\reg$  to
$(\hat \gamma k \delta^{-1}) b\in \cM_1^\reg$.
The final outcome is the following identification:
\be
\cM_\red^\reg = \cN^\reg/\T_n.
\label{P25}\ee
The action  (\ref{P26}) of $\T_n$  on $\cN^\reg$
 is actually a free action. This is a consequence of
the  fact \cite{FG} that
the action of $(K_+(\sigma) \times K_+)/ \Z_{2n}^\diag$ on $\cM_0$ is free.

\medskip\noindent
{\bf Remark 3.1.}
Every element of $\cM_0$ can be mapped into $\cM_1$ by a gauge transformation, which  is
unique up to  residual  gauge transformations acting on $\cM_1$.
It is a useful fact that  \emph{locally},
in a neighbourhood of any fixed element of $\cM_0^\reg$, a well-defined map $f_0$ can be chosen,
\be
f_0\colon \cM_0^\reg \ni g \mapsto g_1\in \cM_1^\reg,
\label{f01}\ee
in such a  manner that the gauge transformed matrix $g_1$  depends
analytically on the local coordinates on the manifold $\cM_0^\reg$.
We next explain this statement.

Let $P^\reg$ denote the manifold of $n\times n$ Hermitian matrices having distinct, positive eigenvalues,
and $G^\reg$ denote the open subset of $\GL(n,\C)$ diffeomorphic to $P^\reg \times \U(n)$ by
the polar decomposition, presented as
\be
\chi = p(\chi) u(\chi),
\qquad
\chi \in G^\reg,\, p(\chi)\in P^\reg,\, u(\chi)\in \U(n).
\label{P27}\ee
Here, $p(\chi)$ and $u(\chi)$ depend analytically on $\chi$.
Let $D^\reg\subset P^\reg$ denote the manifold of real diagonal matrices
$\beta = \diag(\beta_1,\dots, \beta_n)$ satisfying $\beta_1 > \dots > \beta_n >0$.
We recall that  $P^\reg$ is diffeomorphic to $D^\reg \times (U(n)/\T(n))$ by the correspondence
\be
p = \xi(p) \beta(p) \xi(p)^{-1}
\quad\hbox{where} \quad \xi(p) \T(n) \in \U(n)/\T(n)
\label{P28}\ee
with the standard maximal torus $\T(n)< \U(n)$.
Invoking the fact \cite{KN} that $\U(n)$ is a locally trivial bundle over the coset space $\U(n)/\T(n)$,
we see that $\xi(p)\in \U(n)$  in (\ref{P28}) can be \emph{locally}
chosen to be a well-defined, smooth function of $p$.
Now  introduce $\zeta(\chi):= (\det (\xi(p(\chi))^{-2} u(\chi)))^{\frac{1}{2n}}$,
choosing it so as to give a smooth function locally around a fixed $\chi$ at hand.
As the final outcome,   a  locally well-defined map $f_0$ (\ref{f01}) is obtained
as follows:
\be
f_0\colon g \mapsto g_1(g) = g \eta_R(g)^{-1}
\quad \hbox{with}\quad
\eta_R(g)^{-1}=
 \zeta(\chi) \begin{bmatrix}  \xi(p(\chi))   & 0 \\
0  &   u(\chi)^{-1} \xi(p(\chi)) \end{bmatrix}\in K_+,
\label{P29}\ee
 where $\chi:=\chi(g)$ is the upper-right block of $b$ in $g=kb$. Since $\chi$ depends analytically on $g$,
the local choices  guarantee that $g_1(g)$ depends analytically
on the coordinates on $\cM_0^\reg$.

We remark in passing that $\beta_n^2$ resulting from (\ref{P10}) is the smallest eigenvalue of
$\chi \chi^\dagger$, and therefore
$\beta_n$ is not a smooth function on $\cM_0$ at those points where it vanishes.
As we shall see later (from equation (\ref{C12}) and Theorem 5.6),  the assumption (\ref{I13})  excludes this eventuality.

\medskip

In the above we established
the various identifications only at the set-theoretic level.
Although we shall not rely on it technically,
we wish to note
that all above identifications hold
in the category of smooth (and analytic) manifolds as well.
We next prove a lemma, which implies that $\cM_1^\reg$ is an embedded submanifold of $\cM_0$;
itself  known---from \cite{FG}---to be an embedded submanifold of $\cM$.
Utilizing the assumption  (\ref{I13}), it will be shown later that $\cM_1^\reg= \cM_1$.
  Then it follows that $\cM_1 \subset \cM_0$ represents a
reduction of the structure group
(\ref{T34}) of the principal fibre bundle $\cM_0$ over $\cM_\red$ to the subgroup
$(K_+(\sigma) \times \T_{n-1})/\Z_{2n}^\diag$, and $\cN$ is a principal fibre bundle over $\cM_\red$
with structure group
$\T_n$, in the standard sense  \cite{KN}.

\medskip
\noindent{\bf Lemma 3.2.} \emph{Define ${\tilde \cM}_1 \subset \cM_0$ to be the common level set,
at zero value, of the analytic functions $\phi_\xi$ on $\cM_0$ given by
\be
\phi_\xi(g) = \Im \tr (\xi \chi),
\label{P30}\ee
where $\xi$ is any $n\times n$  complex matrix with real diagonal,  and we use  $\chi$ in (\ref{T31}).
Let ${\tilde \cM}_1^\reg \subset {\tilde \cM}_1$ consist of the elements $g=kb$ with $b$ of the form
(\ref{P5}), but  $\beta\in \R^n$ now restricted by  $\beta_i \neq \beta_j$ for $i\neq j$
and $\prod_{i=1}^n \beta_i \neq 0$.
Then the exterior derivatives of the functions $\phi_\xi$ are linearly independent at each point of
${\tilde \cM}_1^\reg$.}

\medskip\noindent{\bf Proof.}
Take an arbitrary $g\in \tilde \cM_1^\reg$ and note that the infinitesimal
gauge transformations by the elements of $\mathrm{Lie}(K_+)$ generate a $(2n-1)n$ dimensional
subspace of the tangent space $T_g \cM_0$, which coincides with the dimension of the
real linear space of the matrices $\xi$.
A general element of $\mathrm{Lie}(K_+)$ is a matrix of the form
\be
 \diag(X,Y),
\qquad
X,Y\in \mathrm{u}(n), \quad \tr(X+Y)=0,
\label{P31}\ee
 and denoting  the induced tangent vector by $V_{(X,Y)}(g)$, we find the derivative
\be
\langle d\phi_\xi(g), V_{(X,Y)}(g) \rangle = \Im \tr\left(\xi (  X \beta - \beta Y )\right), \qquad
\forall g\in \tilde\cM_1^\reg.
\label{P32}\ee
One can easily check that this derivative
vanishes for every $(X,Y)$
if and only if $\xi=0$.
This means that the exterior derivatives $d\phi_\xi(g)$ span a $(2n-1)n$ dimensional subspace
of $T^*_g \cM_0$ at  each $g\in \tilde \cM_1^\reg$, which establishes the claim.
\hfill $\square$
\medskip

The statement of the lemma is non-trivial only if $\tilde{\cM}_1^\reg$ is non-empty, which
turns out to hold. We then also have the non-empty open subset $\tilde \cM_0^\reg$,
which is defined by the condition that the real parts of the diagonal entries of $\chi$ are
pairwise distinct and non-zero. The lemma implies directly that $\tilde \cM_1^\reg$ is an embedded
submanifold of $\tilde \cM_0^\reg$, and hence it is an embedded submanifold of $\cM_0$, too.
Finally, we see that $\cM_1^\reg$ (specified by (\ref{P12})) is
itself an open subset of $\tilde \cM_1^\reg$.   It turns out to be non-empty,
and is therefore also an embedded submanifold of $\cM_0$.

Eventually, we shall obtain the desired model $M$ of $\cM_\red$ as an explicit global cross-section
for the action of $\T_n$ on $\cN = \cN^\reg$.
We shall use Remark 3.1 to show the analyticity of the natural map from
$\cM_0$ onto this cross-section.
This  will enable  us to prove that the construction
gives a model of the symplectic manifold $(\cM_\red, \omega_\red)$.
The procedure is summarized in the following
commutative diagram:

\begin{figure}[h!]
\begin{center}
\begin{tikzpicture}[scale=1.5]
\node (A) at (0,2) {${\mathcal M}_1$};
\node (B) at (2,2) {${\mathcal M}_0$};
\node (F) at (4,2) {$\mathcal M$};
\node (C) at (0,1) {$\mathcal N$};
\node (D) at (0,0) {$M$};
\node (E) at (2,0) {${\mathcal M}_{\red}$};
\path[right hook->] (A) edge node[below]{$\iota_1$}(B);
\path[right hook->] (B) edge node[below]{$\iota_0$}(F);
\path[->,font=\scriptsize,>=angle 90]
(A) edge node[left]{$\pi_1$} (C)
(B) edge node[right]{$$} (E)
(B) edge node[right]{$\psi$} (D)
(B) edge[dashed, bend right] node[above]{$f_0$} (A)
(C) edge node[left]{$\pi_{\mathcal N}$} (D)
(E) edge node[below]{$\Psi$} (D);
\end{tikzpicture}
\end{center}

\caption{Construction of the model $M$ of $\cM_\red$.
The vertical arrows and $\psi$ denote bundle projections; $\iota_1$ and $\iota_0$ are
 embeddings.
  The sets $\cM_0$, $\cM_1$ and $\cN$ are respectively defined in (\ref{defM0}),
 (\ref{P6}) and (\ref{P7}). The arrow $f_0$ represents a locally well-defined gauge transformation (\ref{f01})
 depending  smoothly on $\cM_0$. The map $\pi_1$  is given by (\ref{P7}) and (\ref{P8}).
 The map $\pi_\cN$ denotes the realization of the quotient (\ref{P25})  provided by Proposition 6.4 and Theorem 6.5.}
\label{figure Y}
\end{figure}

\section{A useful characterization of the space $\cN$}
\label{sec:4}

Proposition 4.3 below establishes the equations that determine the image of the map (\ref{P8}),
which can be identified with the space $\cN$ (\ref{P7}). More precisely, we shall proceed
with the help of new variables $(\tilde w, Q, \lambda)$
equivalent to $(w,L, \beta)$. The usefulness of this  characterization
lies in the fact that we will be able to describe
all solution of the constraint equation (\ref{cruc3}) explicitly, and shall rely on this
 to construct the desired model $M$ of $\cM_\red$ (\ref{P25}).

We start by recalling a lemma from \cite{FM}.

\medskip\noindent
{\bf Lemma 4.1.} \emph{The left-handed momentum constraint on $g\in \cM$ defined by (\ref{T30}) is
equivalent to the condition
\be
y^2 g g^\dagger - \frac{1}{2} g g^\dagger (\1_{2n} - I) g g^\dagger =
\frac{1}{2} \alpha^2 (\1_{2n}+ I) + \hat w  \hat w^\dagger,
\label{cruc1}\ee
where $\hat w\in \C^{2n}$ is the fixed vector introduced in (\ref{P3}).}

\medskip\noindent {\bf Proof.}
Irrespective of the constraints, $b_L$ in $g=b_L k_R\in \cM$ can be written as
\be
b_L =
\begin{bmatrix} b_1  & \chi_L \\
0  & b_2  \end{bmatrix},
\label{C2}\ee
and the left-handed constraint requires that $b_2 = y \1_n$
and $b_1 = y^{-1} \sigma$.
By simply spelling it out for $g= b_L k_R$,
  the matrix on the L.H.S. of (\ref{cruc1}) reads explicitly as
 \be
y^2 \begin{bmatrix} b_1b_1^\dagger + \chi_L \chi_L^\dagger   & \chi_L b_2^\dagger \\
 b_2\chi_L^\dagger & b_2 b_2^\dagger \end{bmatrix}
 -
 \begin{bmatrix} \chi_L b_2 b_2^\dagger  \chi_L^\dagger   & \chi_L b_2^\dagger b_2 b_2^\dagger \\
 b_2b_2^\dagger b_2 \chi_L^\dagger & (b_2 b_2^\dagger)^2 \end{bmatrix}.
\label{C3}\ee
The equality of the bottom-right blocks on the two sides of (\ref{cruc1})
is equivalent to $b_2 = y\1_n$.
Then the off-diagonal blocks on both
sides are zero, while  (using (\ref{T29})) the top-left block boils
down to the equality $b_1 b_1^\dagger = y^{-2} \sigma \sigma^\dagger$, which implies the statement.
\hfill $\square$
\medskip

From now on we work on $\cM_1\subset \cM_0$ (\ref{P6}).
Taking any quasi-diagonal $b$, it will prove useful to diagonalize
the positive definite matrix
\be
bb^\dagger = \begin{bmatrix}e^{-2v}\1_n + \beta^2& e^v \beta\\ e^v \beta &e^{2v}\1_n\end{bmatrix}.
\label{C4}\ee
 Let us introduce the real functions $s(t)$ and $c(t)$ by the
formulae\footnote{If $v=0$ then in the limit $t\rightarrow0$ we have $s(0)=c(0)= 1/\sqrt{2}$.}
\be
c(t):= \left[\frac{e^{2t} - e^{2v}}{e^{2t} - e^{-2t}} \right]^\frac{1}{2},
\quad
 s(t):= \left[\frac{e^{2v} - e^{-2t}}{e^{2 t} - e^{-2t}} \right]^\frac{1}{2},
 \qquad \forall t \geq \vert v \vert,
\label{C5}\ee
which imply the identity $c^2(t) + s^2(t)=1$.
Then consider $\lambda \in \R^n$ subject to the condition
\be
\lambda_1 \geq \lambda_2\geq \dots \geq \lambda_n \geq \vert v\vert,
\label{C6}\ee
and define the diagonal matrices
\be
\Lambda (\lambda):= \diag (e^{2\lambda_1},\dots, e^{2\lambda_n}, e^{-2 \lambda_1}, \dots, e^{-2\lambda_n}),
\label{C7}\ee
\be
C(\lambda) = \diag(c(\lambda_1),\dots, c(\lambda_n)),\qquad
S(\lambda) = \diag(s(\lambda_1),\dots, s(\lambda_n)),
\label{C8}\ee
and the matrix
\be
\rho(\lambda) = \begin{bmatrix} C(\lambda)  & S(\lambda) \\
S(\lambda)  & -C(\lambda)  \end{bmatrix}.
\label{C9}\ee
Notice that $\rho(\lambda)$ is real, symmetric  and orthogonal,
\be
\rho(\lambda) = \rho(\lambda)^* =\rho(\lambda)^\dagger = \rho(\lambda)^{-1}.
\label{C10}\ee
Here and throughout the paper, the suffix star on matrices and vectors denotes complex conjugate, and dagger
denotes Hermitian adjoint.

\medskip\noindent
{\bf Lemma 4.2 \cite{FM}.} \emph{For any quasi-diagonal $b$ given by (\ref{P5}),
$b b^\dagger$  can be written as
\be
b b^\dagger = \rho(\lambda) \Lambda(\lambda) \rho(\lambda)^{-1},
\label{C11}\ee
where $\beta$ is related to $\lambda$ according to the one-to-one correspondence
\be
\beta_i= \sqrt{ 2 (\cosh( 2 \lambda_i) - \cosh (2 v))} = 2 \sqrt{\sinh(\lambda_i + v) \sinh(\lambda_i-v)},
 \qquad  i=1,\dots, n.
\label{C12}\ee
}

Now we are ready to formulate the main result of this section.

\medskip\noindent
{\bf Proposition 4.3.} \emph{Take an arbitrary $g = k b \in \cM_1$ (\ref{P6}) for which $\beta$ and $\lambda$
are connected by (\ref{C12}),
and (using $L$ and $w$ from (\ref{P8}))
define $Q\in \U(2n)$ and  $\tilde w\in \C^{2n}$ by
\be
Q:= \rho(\lambda)^\dagger L I \rho(\lambda) = \rho(\lambda)^\dagger k^\dagger I k \rho(\lambda)
\quad\hbox{and}\quad \tilde w:= \rho(\lambda)^\dagger w=  \rho(\lambda)^\dagger k^\dagger \hat w.
\label{C13}\ee
Then the matrix $Q$ and the vector $\tilde w$ satisfy the constraint equation
\be
\Lambda(\lambda) Q \Lambda(\lambda) - \alpha^2 Q = (\Lambda(\lambda)^2 +
\alpha^2 \1_{2n} - 2 y^2 \Lambda(\lambda)) +
2\tilde w  {\tilde w}^\dagger
\label{cruc3}\ee
and the relations
\be
{\tilde w^\dagger}\tilde w=  \alpha^2 (\alpha^{-2n} - 1),
\quad
Q\tilde w = \tilde w.
\label{aux}\ee
Conversely, pick  $\lambda \in \R^n$ verifying (\ref{C6})
and suppose that a matrix   $Q\in \U(2n)$ and a vector $\tilde w\in \C^{2n}$ satisfy (\ref{cruc3})
as well as the
relations (\ref{aux}) and the condition that $Q$ is conjugate to $I$ (\ref{T10}).
Then there exists  $g=kb \in \cM_1$
from which $Q$ and $\tilde w$
can be constructed according to (\ref{C13}), and
such $g$ is unique up to left-multiplication by the elements of the
subgroup $K_+(\hat w)$ (\ref{P3}) of the left-handed gauge group $K_+(\sigma)$. }

\medskip
\noindent
{\bf Proof.}
It is readily checked that the constraint equation (\ref{cruc1}),  together with (\ref{T29}) and
the definitions of $I$ in (\ref{T10}) and $\hat w$ in (\ref{P3}),
implies (\ref{cruc3}) and (\ref{aux}) for $Q$
and $\tilde w$ defined by (\ref{C13}).

In order to prove the converse, which
gives the reconstruction of $g\in \cM_1$ from $\lambda$, $Q$ and $\tilde w$,
we start by noting that if $Q\in \U(2n)$ is conjugate to $I$ (\ref{T10}),
then we can a find an element $\kappa\in K$ for which
\be
\rho Q \rho^{-1} = \kappa^\dagger I \kappa, \quad\hbox{where}\quad \rho := \rho(\lambda).
\label{C16}\ee
Next, we observe that the auxiliary condition $Q \tilde w = \tilde w$ is equivalent to
\be
I \kappa \rho \tilde w = \kappa \rho\tilde w.
\label{C17}\ee
By using (\ref{C17})  and the property that
  ${\tilde w^\dagger}\tilde w= \alpha^2 (\alpha^{-2n} - 1)$, we see that there
exists an element $k_+ \in K_+$ for which
\be
k_+ \kappa \rho \tilde w =  \hat w.
\label{C18}\ee
Let us now define $g=kb$ by using
\be
k:= k_+ \kappa
\label{C19}\ee
together with the quasi-diagonal $b$ associated to $\lambda$ via (\ref{C12}).
Then routine manipulations show that  equation (\ref{cruc3}) implies for $g$ the left-handed momentum
map constraint (\ref{cruc1}).

 Now let us inspect the ambiguity in the above constructed $k$, and thus in $g$.
If $\kappa'$ and $k_+'$ represent another choice in the above equalities, then we have
\be
\kappa'= \eta_+ \kappa
\quad
\hbox{for some}\quad \eta_+\in K_+,
\label{C20}\ee
and thus
\be
k_+ \kappa \rho \tilde w = k_+'\kappa'  \rho \tilde w = k_+' \eta_+ \kappa \rho \tilde w =
 \hat w.
\label{C21}\ee
Therefore
\be
k_+^\dagger \hat w = (k_+' \eta_+)^\dagger \hat w,
\label{C22}\ee
and hence
\be
k_+' \eta_+ =  \hat \eta_L k_+
\quad\hbox{for some}\quad \hat \eta_L \in K_+(\hat w).
\label{C23}\ee
This entails that
\be
k'= k_+' \kappa' = k_+' \eta_+ \kappa= \eta_L k_+ \kappa = \hat \eta_L k
\quad\hbox{and}\quad
g' = k' b = \hat \eta_L g,
\label{C24}\ee
that is, $k$ and $g$ are unique  up to left-multiplication by the isotropy subgroup of the vector
$\hat w$ in $K_+$.
\hfill $\square$
\medskip

\noindent
{\bf Definition 4.4.}
Let us call a triple $(\tilde w, Q,\lambda) \in \C^{2n}\times \C^{2n\times 2n} \times \R^n$ \emph{admissible} if
$\lambda$ satisfies (\ref{C6}),
the constraint equation (\ref{cruc3}) holds,   $Q$ is  unitary, conjugate to $I$ (\ref{T10}),  and
the auxiliary conditions (\ref{aux}) are met.
Denote by $\cN(\lambda)$ the set of admissible triples associated with fixed $\lambda$,
and let $\cM_1(\lambda)$  stand for the subset of $\cM_1$ corresponding, by Proposition 4.3,
to the admissible triples
with fixed $\lambda$. Moreover, denote by $\cM_0(\lambda)$ the subset of $\cM_0$ whose elements
can be gauge transformed into $\cM_1(\lambda)$.
Finally, denote by  $\cD(u,v,\mu)$ the set of the admissible $\lambda$ variables, i.e.,
those that appear in admissible triples.

\medskip

It is clear from the relations (\ref{C12}) and (\ref{C13})  that the
 triple $( \tilde w, Q,\lambda)$ is equivalent
to the triple $( w, L,\beta)$ (in the obvious sense that one can be expressed in terms of the other).
By using this equivalence, and Proposition 4.3,
 we identify $\cN(\lambda)$ as defined above with
the image of the map (\ref{P8}), with $\beta$ taking the value (\ref{C12}).

We now elaborate the gauge transformation properties of the variables $( \tilde w,Q, \lambda)$.
For this, we
  note first of all that
if a triple  $( \tilde w,Q, \lambda)$ is admissible,  then  so is
 $( \gamma^{-1} \tilde w, Q, \lambda)$ for any $\gamma\in \U(1)$.
This reflects the gauge freedom
whereby the elements $g\in \cM_1$ are transformed as $g \mapsto \hat \gamma g$
with $\hat \gamma \in \T_1$ (\ref{P4}).
The set  $\cM_1(\lambda)$ is also
mapped to itself by the right-handed gauge transformations generated by those
$\eta_R = \diag(\eta_R(1), \eta_R(2)) \in K_+$
 for which
 \be
\eta_R(1)\diag(\beta_1(\lambda),\dots, \beta_n(\lambda)) \eta_R(2)^{-1} =
\diag(\beta_1(\lambda), \dots, \beta_n(\lambda)).
\label{C25}\ee
 We denote the corresponding
subgroup of the right-handed gauge group $K_+$ by
 $K_+(\lambda)$.
 Using this and the relations (\ref{P2}) and (\ref{P7}),
it is readily seen that Proposition 4.3 gives rise to the following
natural identifications:
\be
\cM_\red(\lambda) :=  K_+(\sigma)\backslash \cM_0(\lambda)/K_+= K_+(\sigma)\backslash
 \cM_1(\lambda)/   K_+(\lambda) =\T_1\backslash \cN(\lambda)/ K_+(\lambda),
\label{C26}\ee
where the last quotient refers to the gauge transformations
\be
\cN(\lambda)\ni (\tilde w, Q, \lambda) \mapsto ( \gamma^{-1} \eta_R \tilde w, \eta_R Q \eta_R^{-1}, \lambda ),
\qquad
\forall (\hat \gamma, \eta_R)\in \T_1\times K_+(\lambda).
\label{C27}\ee
In the regular case (\ref{P12}), we have
\be
\lambda_1 > \lambda_2 > \dots > \lambda_n > \vert v \vert,
\label{C28}\ee
and $K_+(\lambda) =\T_{n-1}$.
Then, the transformations (\ref{C27})  yield
the gauge action of $\T_n$ (\ref{P15}):
 \be
 ( \tilde w, Q,\lambda) \mapsto ( \tau\tilde w, \tau Q \tau^{-1}, \lambda),\qquad
\forall \tau \in \T_n,
\label{C29}\ee
which is completely equivalent to (\ref{P26}) via the definitions in (\ref{C13}).

In order to construct the desired model of $\cM_\red$, we need to describe all admissible
triples  $(\tilde w, Q,\lambda)$.
A crucial part of this problem is  to find the admissible $\lambda$,  which parametrize the eigenvalues
of $bb^\dagger$ for $g=k b\in \cM_0$.  These eigenvalues, and thus also the components
of $\lambda$, can be viewed as continuous functions on $\cM_0$,  and we are
looking for the range of the corresponding map, $\cL$,
\be
 \qquad \cD(u,v,\mu) = \cL(\cM_0) \quad\hbox{with}\quad \cL\colon   g \mapsto \lambda.
\label{C30}\ee
In the following section, we shall  describe  $\cD(u,v,\mu)$ and the corresponding solutions
of (\ref{cruc3}) explicitly. See  Theorem 5.6 for the result.

We can explain at this point why an open subset of the reduced phase space
can be parametrized by the $\lambda_i$ together with $n$ angular variables;
which appear in (\ref{I8}).
To this end, let us take an arbitrary element
\be
e^{\ri \xi} \equiv \diag(e^{\ri \xi_1},\dots, e^{\ri \xi_{2n}})
\label{C31}\ee
from the torus $\T^{2n}$, and notice that if $( \tilde w, Q, \lambda)$ is admissible,
then so is
\be
( e^{\ri \xi} \tilde w, e^{\ri \xi} Q e^{-\ri \xi},\lambda),\qquad \forall e^{\ri \xi}\in \T^{2n}.
\label{C32}\ee
  Indeed, the conditions described in Proposition 4.3 are respected by these transformations.
In addition to the gauge transformations by $\tau\in \T_n$ in (\ref{C29}), these $\T^{2n}$ transformations
involve $n$ arbitrary angles, which parametrize $\T^{2n}/\T_n$.
It is clear that, for generic $\lambda$, equation (\ref{cruc3})
permits the expression of $Q$ in terms of
$\lambda$ and $\tilde w$. Moreover, we shall see shortly that the  $\vert \tilde w_a\vert $
can  be expressed in terms of $\lambda$, and generically none of
them vanish.
This implies that generically the elements of
$\cN(\lambda)/\T_n$ can indeed be parametrized by $n$-angles.

\medskip\noindent
{\bf Remark 4.5.}
We know that the $\T_n$ action on $\cN(\lambda)$ is free, and shall also confirm this explicitly later.
Moreover, it will turn out that the $\T^{2n}$ action, sending $(\tilde w,Q,\lambda)$ to (\ref{C32}),
is transitive on $\cN(\lambda)$; and is also free
except for a certain lower dimensional subset of the admissible $\lambda$ values.

\section{Solution of the constraints}
\label{sec:5}

Locally, the general solution of the constraint equation (\ref{cruc3}) was already
found in \cite{FM}.
Here, `locally'  means that the form of the domain of the $\lambda$-variables
was not established.
In this section, we shall prove that $\cD(u,v,\mu)$ (\ref{C30}) is the
closure of $\cD_+$  in (\ref{I9}), as was anticipated in \cite{FM}.
Moreover, we shall describe all admissible triples forming $\cN$ (\ref{P7}) explicitly.
When combined with the local results of \cite{FM}, this yields a model
of the reduced system coming from the Abelian Poisson algebra $\fH^1$ (\ref{fH1})
restricted
to a dense open submanifold, and will
permit us to derive the desired global model  $ M$ of $\cM_\red$ in Section 6.

For technical reasons that will become clear shortly, we initially work on a certain
dense open subset of $\cM_0$. To define this subset, let us consider the following
symmetric polynomials
in $2n$ indeterminates:
\be
 p_1(\Lambda) = \prod_{k\neq \ell}^{2n} (\Lambda_k - \Lambda_\ell) (\Lambda_k \Lambda_\ell - \alpha^2),
 \label{F1}\ee
 and
 \be
  p_2(\Lambda) = \prod_{k=1}^{2n}  (\Lambda_k - \alpha)( y^2 \Lambda_k  - \alpha^2)(\Lambda_k - y^2) (\Lambda_k -  x^{2}).
 \label{F2}\ee
Since  $\cM_0$ (\ref{defM0}) is a joint level surface of independent
 analytic functions on $\cM$, it
  is an analytic submanifold of $\cM$, and
 thus
 we obtain analytic functions on $\cM_0$ if we substitute
 the eigenvalues $\Lambda_k(g)$ of $ g g^\dagger = k bb^\dagger k^{-1}$ into the above polynomials.
 This  follows since, being symmetric polynomials in the eigenvalues,
 the $p_i(\Lambda(g))$
 can be expressed as polynomials in the coefficients of the characteristic polynomial of $g g^\dagger$.
We know that $\cM_0$ is connected and, as explained in Remark 5.1,
 can also conclude that
\be
p(g):=  p_1(\Lambda(g))  p_2(\Lambda(g))
\label{F3}\ee
 does
not vanish identically on $\cM_0$.
By analyticity, this implies that
\be
\cM_0^{\sreg}:= \{ g\in\cM_0\mid  p(g)\neq 0\}
\label{F4}\ee
is a \emph{dense} open subset of $\cM_0$.
We call its elements \emph{strongly regular}. We shall apply the same adjective to the
$\lambda$-values for which (using (\ref{C7}))  $ p(\Lambda(\lambda)) \neq 0$,
and call also strongly regular the corresponding admissible triples $( \tilde w, Q, \lambda)$,
whose set is denoted $\cN^\sreg$.
The admissible strongly regular $\lambda$-values form the dense subset
\be
\cD(u,v,\mu)^\sreg = \cL(\cM_0^\sreg) \subset \cD(u,v,\mu).
\label{F5}\ee

\medskip
\noindent
{\bf Remark 5.1.}
Let us recall  from \cite{FG} that the reductions of the Hamiltonians
\be
\hat \cH_j(g) = \frac{1}{2}\tr\!\left((b b^\dagger)^j\right), \qquad j=1,\dots, n,
\label{F6}\ee
provide a Liouville integrable system on the $2n$-dimensional reduced phase space $\cM_\red$.
These reduced Hamiltonians can be expressed in terms of the $\lambda_i$ ($i=1,\dots, n$) as
\be
\hat \cH_j^\red =  \sum_{i=1}^n \cosh (2j \lambda_i).
\label{F7}\ee
Their functional independence implies that the range of the $\lambda$-variables
must contain an open subset of $\R^n$. It follows from this that $\cM_0^{\sreg}$
 cannot be empty.

\medskip

Focusing  on $\cN^\sreg$,
we introduce the $2n \times 2n$ diagonal matrices
\be
\cW:= \diag(\tilde w_1,\dots, \tilde w_{2n}),\qquad
D_{l m} = \frac{ \Lambda_l^2 + \alpha^2 - 2y^2 \Lambda_l}{\Lambda_l^2 - \alpha^2} \delta_{lm},
\label{F8}\ee
and the Cauchy-like matrix $C$,
\be
C_{lm} := \frac{1}{\Lambda_l \Lambda_m - \alpha^2}.
\label{F9}\ee
The denominators do not vanish since $\lambda$ is strongly regular.
The constraint equation (\ref{cruc3}) leads to the
following formula for the matrix $Q$:
\be
Q = D  + 2\cW C {\cW^\dagger}.
\label{F10}\ee
Since $Q$ is conjugate to $I$ (\ref{T10}),  $Q^2 = \1_{2n}$ holds, and this translates into
\be
D^2 + 2\cW D C \cW^\dagger + 2\cW C D \cW^\dagger + 4\cW C (\cW \cW^\dagger) C \cW^\dagger = \1_{2n}.
\label{F11}\ee
Let us observe that the matrix $\cW$ is
invertible whenever $\lambda$ is strongly regular.
Indeed, if some component $\tilde w_a=0$, then (\ref{F11}) yields $D_a^2=1$,
which is excluded by strong regularity.

Next, we substitute (\ref{F10}) into the equation
$Q \tilde w = \tilde w$  in (\ref{aux}), which gives
\be
D_{jj} \tilde w_j + 2\tilde w_j \sum_{m=1}^{2n} C_{jm} \vert \tilde w_m \vert^2 = \tilde w_j,
\qquad
\forall j=1,\dots, 2n.
\label{F12}\ee
Dividing by $\tilde w_j$ produces the  formula
\be
\vert \tilde w_j \vert^2  = \frac{1}{2}
\sum_{l=1}^{2n} (C^{-1}(\lambda))_{jl} ( 1 - D(\lambda)_{ll}),
\label{simple}\ee
where $C^{-1}$ is the inverse of the matrix $C$ (\ref{F9}) and we took into account (\ref{C7}).
This expresses the moduli $\vert \tilde w_j \vert$ as functions of $\lambda$.

Using the parameter $\mu$ instead of $\alpha= e^{-\mu}$,
define the $2n$ functions
\be
\ba
F_a(\lambda)&= \prod_{\substack{i=1\\(i\neq a)}}^{n}
\left(\frac{\sinh(\lambda_a+\lambda_i+\mu)\sinh(\lambda_a-\lambda_i+\mu)}
{{\sinh(\lambda_a-\lambda_i)}\sinh(\lambda_a+\lambda_i)}\right),
\quad
1\leq a \leq n, \\
F_{n+a}(\lambda)&=
 \prod_{\substack{i=1\\(i\neq a)}}^{n}
\left(\frac{\sinh(\lambda_a+\lambda_i-\mu)\sinh(\lambda_a-\lambda_i-\mu)}
{{\sinh(\lambda_a-\lambda_i)}\sinh(\lambda_a+\lambda_i)}\right),
\ea
\label{F14}\ee
as well as the functions
\be
\ba
\cF_a(\lambda) &=  e^{-\mu}\left(e^{2 \lambda_a} - y^2\right) \frac{\sinh(\mu)}{\sinh(2\lambda_a)} F_a(\lambda),
\quad 1\leq a\leq n, \\
\cF_{n+a} (\lambda)&=  e^{-\mu}\left(y^2 - e^{-2 \lambda_a}\right) \frac{\sinh(\mu)}{\sinh(2\lambda_a)} F_{n+a}(\lambda).
\ea
\label{F15}\ee

\medskip\noindent
{\bf Proposition 5.2.}
\emph{
The moduli of $\tilde w_j(g)$ defined by (\ref{C13}) are gauge invariant functions of
 $g=kb\in\cM_1^\reg$ and depend only
on  $\lambda$ that parametrizes the eigenvalues of $bb^\dagger$ according to (\ref{C7}) and (\ref{C11}).
Explicitly, these functions are given by the relation
\be
\vert \tilde w_j(g) \vert^2 = \cF_j(\lambda), \qquad j=1,\dots, 2n,
\label{*}\ee
with the functions $\cF_j$ (\ref{F15}).
The component $Q$ in any admissible strongly regular triple $( \tilde w, Q,\lambda)\in \cN^\sreg$
can be written as (\ref{F10}), where the phases of the entries of
$\tilde w\in \C^{2n}$ can be chosen arbitrarily.}

\medskip\noindent
{\bf Proof.} For a strongly regular admissible $\lambda$, the formula (\ref{*}) is a
reformulation  \cite{FM}
of (\ref{simple}).
It remains valid on the whole of $\cM_1^\reg$, since the functions on the two sides
of (\ref{*}) are gauge invariant continuous functions on $\cM_1^\reg$, and
$\cM_1^\sreg$ is a dense subset of $\cM_1^\reg$ (in consequence of the density
of $\cM_0^\sreg$ in $\cM_0$).
In the strongly regular case, the formula (\ref{F10}) for $Q$ was derived above.
The phases of $\tilde w_j$ can take arbitrary values, because one can
use arbitrary $e^{\ri \xi}\in \T^{2n}$  in equation (\ref{C32}).
\hfill $\square$

\medskip

The definitions guarantee the positivity of $\vert \tilde w_j\vert(\lambda)$ for every
$\lambda\in \cD(u,v,\mu)^\sreg$
(see below (\ref{F11})).
Thus, the explicit formula (\ref{*}) leads to a necessary condition on $\lambda$ to belong to the
(still unknown) set $\cD(u,v,\mu)^\sreg$.
Indeed, our aim below is to identify  the `maximal domain' on which the functions $\cF_j$
\emph{as given by the formula} (\ref{F15}) are positive.
More precisely,
we are interested in the set
\be
\cD_+(u,v,\mu):= \{ \lambda\in \R^n\mid \lambda_1> \lambda_2 > \dots > \lambda_n > \vert v \vert,\,\,
\cF_j(\lambda)>0,\,\,
\forall j=1,\dots, 2n \}.
\label{F17}\ee
We stress that in this definition $\lambda$ is  \emph{not} assumed to be admissible or strongly regular;
the formula (\ref{F15}) is used to define $\cF_j(\lambda)$ for the $\lambda$ that occur.
Next, we shall give the elements of $\cD_+(u,v,\mu)$ explicitly.
After that, we shall prove that $\cD(u,v,\mu)$ (\ref{C30}) is the closure of $\cD_+(u,v,\mu)$.
Our notation anticipates that the definition (\ref{F17}) turns out to give the set (\ref{I9}).

\medskip\noindent
{\bf Proposition 5.3.}
\emph{The set $\cD_+(u,v,\mu)$ defined by (\ref{F17}) can be described explicitly as
\be
\cD_+(u,v,\mu)= \{ \lambda\in \R^n\mid \lambda_n > \operatorname{max}(\vert u \vert, \vert v \vert),\,\,
\lambda_i -\lambda_{i+1} > \mu,\, \forall  i=1,\dots, n-1\}.
\label{F18}\ee
}

\noindent
{\bf Proof.}
It is straightforward to check that if $\lambda\in \R^n$ verifies
\be
\lambda_n > \operatorname{max}(\vert u \vert, \vert v \vert),\,\,
\lambda_i -\lambda_{i+1} > \mu,\, \forall  i=1,\dots, n-1,
\label{F19}\ee
then $\cF_j(\lambda)>0$, and actually also $F_j(\lambda)>0$, for all $j=1,\dots, 2n$.

To prove the converse, suppose that $\lambda$ meets the requirements imposed in (\ref{F17}), and
that it also satisfies
\be
\lambda_n > - u.
\label{F20}\ee
This latter assumption holds automatically for $\vert v \vert > \vert u\vert $, and also when   $\vert u \vert > \vert v \vert$ if $u>0$.
It follows from (\ref{F20}) that
\be
(e^{2 \lambda_a} - y^2)= (e^{2 \lambda_a} - e^{-2 u}) >0,
 \label{F21}\ee
 and hence the positivity of $\cF_a(\lambda)$ implies
\be
F_a(\lambda) >0, \quad \forall a=1,\dots, n.
\label{F22}\ee
We note that $F_1(\lambda) >0$ holds  as a consequence of $\lambda_1>\lambda_2> \dots >\lambda_n > \vert v\vert$.
Then we look at $F_2$ and find that $F_2(\lambda) >0$ forces
$\lambda_1 - \lambda_2 >\mu$.
Next we inspect $F_3$, and wish to show  that its positivity implies
$\lambda_2 - \lambda_{3} > \mu$.
For this, we notice that the only factors in $F_3$ that are not manifestly positive are those in the product
\be
\frac{\sinh(\lambda_3 - \lambda_1 + \mu)}{\sinh(\lambda_3 - \lambda_1)}\frac{   \sinh(\lambda_3-\lambda_2 +\mu) }{\sinh(\lambda_3 - \lambda_2)}.
\label{F23}\ee
We  recast this product slightly as
\be
\frac{\sinh(\lambda_1 - \lambda_2 - \mu + (\lambda_2 - \lambda_3))}{\sinh(\lambda_1 - \lambda_3)} \frac{\sinh(\lambda_2 - \lambda_3 - \mu)}{\sinh(\lambda_2 - \lambda_3)},
\label{F24}\ee
and since we already know that $\lambda_1 - \lambda_2 > \mu$, we see that each factor is positive
except possibly $\sinh(\lambda_2 - \lambda_3 -\mu)$. Thus the positivity of $F_3(\lambda)$ leads to
$\lambda_2 - \lambda_3 > \mu$.
We go on in this manner and find that  the positivity of all
\be
F_1(\lambda), F_2(\lambda),\ldots, F_a(\lambda)
\label{F25}\ee
implies (actually is equivalent to)
\be
\lambda_i - \lambda_{i+1} >\mu,
\quad
\forall i=1,\ldots, a-1.
\label{F26}\ee
This holds for each  $a=2,\dots, n$.

We now observe that if $\lambda_i - \lambda_{i+1} >\mu$ for all $i$, then $F_{n+a}(\lambda)>0$ is valid for all $a=1,\dots, n$ as well.
Therefore the positivity of $\cF_{2n}(\lambda)$ requires that
\be
(e^{-2u} - e^{-2\lambda_n})>0,
\label{F27}\ee
which in the case $u>0$ enforces  that $\lambda_n > \vert u\vert $.

At this stage, the proof is complete whenever (\ref{F20}) is guaranteed.  Therefore,
it only  remains to show that $\lambda_n > \vert u \vert $ must
hold also when $\vert u \vert > \vert v \vert$ and $u<0$.
This follows from  Lemma 5.4 below.
\hfill $\square$

\medskip
\noindent
{\bf Lemma 5.4.}
\emph{If $u<0$, then there does not exist any
$\lambda\in \R^n$, $\lambda_1 > \lambda_2 > \dots > \lambda_n >0$ for which $\lambda_n < \vert u \vert $
and the expressions (\ref{F15}) satisfy
 $\cF_m(\lambda) >0$ for all $m=1,\dots,2n$.}

\medskip\noindent
{\bf Proof.}
If $\lambda_n < \vert u\vert$ and $\cF_1(\lambda) >0$ by (\ref{F15}), then there exists a smallest index
$1 < k \leq n$ such that $\lambda_{k-1} > \vert u \vert$ but $\lambda_k < \vert u\vert$.
This follows since $\lambda_1$ must be larger than
$\vert u\vert$, otherwise $\cF_1(\lambda)>0$ cannot hold.
The positivity of $\cF_m(\lambda)$ for all $m$ then requires
\be
F_1(\lambda)>0,\dots, F_{k-1}(\lambda)>0,\,\, F_{k}(\lambda)<0,\dots, F_n(\lambda)<0,\,\,
F_{n+1}(\lambda) >0,\dots, F_{2n}(\lambda)>0.
\label{F28}\ee
Let us now suppose that
\be
2\leq k \leq n-1, \quad (n>2).
\label{F29}\ee
We find that the positivity of $F_1,\dots, F_{k-1}$ is equivalent to
the $(k-2)$ conditions
\be
\lambda_1 - \lambda_2 > \mu,\dots, \lambda_{k-2} - \lambda_{k-1} > \mu.
\label{F30}\ee
In particular, these conditions are empty  for $k=2$.
Then the negativity of $F_k$ leads to the condition
\be
\lambda_{k-1} - \lambda_k < \mu.
\label{F31}\ee
Moreover,  the negativity of $F_{k+1},\dots, F_n$  leads to the conditions
\be
\lambda_{k} - \lambda_{k+1} <\mu, \dots, \lambda_{n-1} - \lambda_n < \mu
\label{F32}\ee
together with
\be
\lambda_{k-1} - \lambda_{k+1} >\mu,\dots, \lambda_{n-2} - \lambda_n >\mu.
\label{F33}\ee
But then we find that the above inequalities imply
\be
F_{n+ k -1}(\lambda) < 0.
\label{F34}\ee
We here used that $\lambda_{k-1} > \mu$, which follows from the above.

We have proved  that $\lambda$ satisfying our conditions does not exist
if $2\leq k \leq n-1$.
It remains to consider the case
$k=n$,
when we must have $F_n(\lambda)<0$, but all
the other $F_k$ must be positive. Inspecting these functions for $k=2,\dots, n-1$ we find
$\lambda_i - \lambda_{i+1} > \mu$ for $i=1,\dots, n-2$ and from $F_n(\lambda)<0$ we find $\lambda_{n-1} - \lambda_n < \mu$.
Then one can check that $F_{n+1},\dots, F_{2n-2}$ are  positive, while
the positivity of $F_{2n-1}(\lambda)$ requires $\lambda_{n-1} + \lambda_n < \mu$. The inequalities derived
so far entail that $F_{2n}(\lambda) < 0$, and thus $\lambda$ with the required properties does not
exist in the $k=n$ case either.

In the above  it was  assumed that $n>2$, but the arguments are easily adapted to cover the $n=2$
case, too.
\hfill $\square$
\medskip

 We see from Proposition 5.3 that the sets given by  (\ref{F5}) and (\ref{F18}) satisfy
\be
\cD(u,v,\mu)^\sreg \subseteq \cD_+(u,v,\mu).
\label{F35}\ee
Since $\cD(u,v,\mu)^\sreg$ is a dense subset of the set  $\cD(u,v,\mu)$ of admissible $\lambda$-values, we
obtain
\be
\cD(u,v,\mu) \subseteq \overline{ \cD_+(u,v,\mu)},
\label{F36}\ee
where
\be
\overline{\cD_+(u,v,\mu)} = \{ \lambda \in \R^n\mid \lambda_n \geq  \operatorname{max}(\vert u\vert, \vert v \vert),\,\,
\lambda_i -\lambda_{i+1} \geq \mu,\, \forall  i=1,\dots, n-1\}
\label{F37}\ee
is the closure of $\cD_+(u,v,\mu)$.
 We shall shortly  demonstrate that in (\ref{F36}) equality holds.

Employing the notation (\ref{C31}), let us take an arbitrary element $e^{\ri \xi}\in \T^{2n}$
and consider, for $l,m=1,\dots,2n$, the formulae
\be
Q_{lm}(\lambda, e^{\ri \xi}) =  D_{lm}(\lambda) + 2\tilde w_l(\lambda,e^{\ri \xi}) C_{lm}(\lambda) {\tilde w_m}^*
(\lambda, e^{\ri\xi}),
\qquad
\tilde w_{l}(\lambda, e^{\ri \xi}) = e^{\ri \xi_l} \sqrt{\cF_l(\lambda)},
\label{F40}\ee
where non-negative square roots are used for all $\lambda \in \overline{\cD_+(u,v,\mu)}$.
The matrix element $Q_{lm}$ shows an apparent singularity
at the $\lambda$-values for which the denominator in $C_{lm}(\lambda)$ (\ref{F9}) becomes zero.
However,
all those `poles' cancel either against zeros of
$\sqrt{\cF_l(\lambda) \cF_m(\lambda)}$ or against a corresponding pole of $D_{lm}(\lambda)$.\\

\medskip\noindent
{\bf Lemma 5.5.} \emph{The formulae (\ref{F40}) for $Q_{lm}$ and $\tilde w_l$ yield unique continuous functions on
the domain
 $ \overline{ \cD_+(u,v,\mu)} \times \T^{2n}$, which are analytic on the
 interior $\cD_+(u,v,\mu)\times \T^{2n}$. The components of $\rho(\lambda)$ (\ref{C9}) and
 $\beta(\lambda)$ (\ref{C12}) are also analytic on  $\cD_+(u,v,\mu)$ and continuous on its closure.
 }

 \medskip
 \noindent {\bf Proof.}
For any fixed $j=1,\dots, n-1$, the matrix element
\be
C_{j+1, n+j}(\lambda) = - \frac{1}{2}
\frac{e^{\mu + \lambda_j - \lambda_{j+1}}}{\sinh(\lambda_j -\lambda_{j+1} -\mu)},
\label{F41}\ee
becomes infinite as $\lambda_j - \lambda_{j+1} -\mu$ tends to zero.
This pole is cancelled by the corresponding zero of
\be
\sqrt{\cF_{j+1}(\lambda) \cF_{n+j}(\lambda)} = \sinh(\lambda_j - \lambda_{j+1} - \mu) f_{j+1, n+j}(\lambda),
\label{F42}\ee
where $f_{j+1, n+j}(\lambda)$ remains finite as $\lambda$ approaches the pole.

The only other source of potential singularity of $Q_{lm}$ (\ref{F40}) is the vanishing of the denominators of
$D_{2n,2n}$ (\ref{F8}) and $C_{2n,2n}$ (\ref{F9}) as $\lambda_n$ tends to $\mu/2$.
This may be excluded  by the form of $\cD_+(u,v,\mu)$, but when it is not excluded then
one can check  easily that these poles cancel against each other.
The continuity of the resulting functions on  $\overline{ \cD_+(u,v,\mu)} \times \T^{2n}$ and their
analyticity on the interior also follow immediately from their explicit formulae.
The statements regarding $\rho(\lambda)$ and $\beta(\lambda)$ are plainly true.
$ \hfill\square$

\medskip

The following theorem summarizes one of our  main results.

\medskip\noindent
{\bf Theorem 5.6.}
\emph{The set of admissible triples $(\tilde w, Q, \lambda)$, which according to Proposition 4.3
is in bijective correspondence
with the set $\cN$ (\ref{P7}), is formed precisely by the triples $(\tilde w, Q,\lambda)$
given explicitly
by Lemma 5.5.
Consequently, the image $\cD(u,v,\mu)$ of the `eigenvalue map' $\cL$ (\ref{C30}) equals the closure
of $\cD_+(u,v,\mu)$, given by (\ref{F37}).
The dense open submanifold of the reduced phase space defined by
\be
\cM_\red^+ := K_+(\sigma) \backslash \cL^{-1}(\cD_+(u,v,\mu))/K_+
\label{F43}\ee
is in bijective correspondence with set of admissible triples given by Lemma 5.5 using
$\lambda\in \cD_+(u,v,\mu)$ and $e^{\ri \xi}$ taking the form
\be
(e^{\ri \xi_1},\dots, e^{\ri \xi_n}, e^{\ri \xi_{n+1}},\dots, e^{\ri \xi_{2n}}) =
(e^{\ri \theta_1},\dots, e^{\ri \theta_n}, 1,\dots, 1)
\quad\hbox{with}\quad e^{\ri \theta}\in \T^n.
\label{F44}\ee
This yields a symplectomorphism  between $\cM_\red^+$ equipped with the restriction of $\omega_\red$
and the product manifold  $ \cD_+(u,v,\mu) \times \T^n$ equipped with the symplectic
form
$ \sum_{j=1}^n d \theta_j \wedge d \lambda_j$.
}

\medskip
\noindent
{\bf Proof.}
In what follows, we first show that all triples
given by Lemma 5.5 are admissible, that is, they represent elements on
$\cN$.  In particular\footnote{From now on we drop $u,v,\mu$
from $\cD(u,v,\mu)$, $\cD_+(u,v,\mu)$ and $\cD^\sreg(u,v,\mu)$.},  $\cD$ defined in (\ref{C30}) is the closure of
$\cD_+$ in (\ref{F18}).
Then we apply a density argument to  demonstrate that the admissible triples of Lemma 5.5 exhaust $\cN$.
Finally, we  explain the statement about the model of the subset $\cM_\red^+$ of $\cM_\red$.

We have seen that  for any $\lambda \in \cD^\sreg\subset \cD$ every admissible triple
$(\tilde w,Q,\lambda)$ is of the form
(\ref{F40}),
and we also know that $\cD^\sreg$ is a non-empty open
subset
of $\cD_+$. By noting that the triple (\ref{C32})
is admissible whenever $(\tilde w,Q,\lambda)$ is admissible, we conclude that the conditions on admissible triples
formulated in Definition 4.4   are satisfied by the triples given by  (\ref{F40})  with
$(\lambda, e^{\ri \xi})$ taken from  the open subset
$\cD^\sreg \times \T^{2n} \subset \cD_+ \times \T^{2n}$.
Because these conditions require the vanishing of analytic functions, they must then hold on the connected
open set $\cD_+ \times \T^{2n}$, and by continuity on its closure as well. Thus, we have proved that all triples
given by Lemma 5.5 are admissible. On account of (\ref{F36}),
 this implies that $\cD = \overline{\cD_+}$.

 We now show that Lemma 5.5 gives \emph{all} admissible triples.
To this end, let us choose an admissible triple, denoted $( \tilde w_\sharp, Q_\sharp, \lambda_\sharp)$, for which
$\lambda_\sharp\in (\cD \setminus \cD^\sreg)$. This corresponds by equation
 (\ref{C13}) to some element $g_{1\sharp}\in \cM _1$, which
is obtained by a right-handed gauge transformation from  some element $g_\sharp\in \cM_0$.
We fix $g_{1\sharp}$ and $g_\sharp$.
We can find a sequence  $g(j)\in \cM_0^\sreg$ that converges to $g_\sharp$, because
$\cM_0^\sreg$ is a dense subset of $\cM_0$.
It is easy to see that the sequence  $g(j)$ can be gauge transformed into a sequence $g_1(j)\in \cM_1$ (\ref{P6})
that \emph{converges to}  $g_{1\sharp}$.
(This follows from the continuous dependence on $g$ of the eigenvalues  $\beta_i^2$ of $\chi \chi^\dagger$,
where $\chi$
is the top-right block  of $b$ from $g=kb\in \cM_0$.)
The convergent sequence  $g_1(j)\in \cM_1$
corresponds by equation (\ref{C13}) to a sequence $(\tilde w(j), Q(j), \lambda(j))$ of strongly regular
admissible triples
that converges to $( \tilde w_\sharp, Q_\sharp, \lambda_\sharp)$.
Then, as for any $\lambda \in \cD^\sreg$ every admissible triple is of the form (\ref{F40}) , we obtain
a sequence
 $(\lambda(j), e^{\ri \xi}(j))\in \cD^\sreg \times \T^{2n}$ that obeys
 \be
 \lim_{j\to \infty} \left(\tilde w\!\left(\lambda(j), e^{\ri \xi}(j)\right)\,,\, Q\!\left(\lambda(j), e^{\ri \xi}(j)\right)\,, \,\lambda(j)\right) =
 \left(\tilde w_\sharp\,,\, Q_\sharp\,,\, \lambda_\sharp\right).
 \label{F46}\ee
By the compactness of $\T^{2n}$, possibly going to a subsequence, we can assume that $e^{\ri \xi}(j)$ converges
to some $e^{\ri \xi_\sharp}$.
By the continuous dependence of the triple in Lemma 5.5 on $(\lambda, e^{\ri \xi})$, it finally follows that
\be
\left(\tilde w_\sharp\,,\, Q_\sharp\,, \,\lambda_\sharp\right) = \left(\tilde w\!\left(\lambda_\sharp, e^{\ri \xi_\sharp}\right)\,, \, Q\!\left(\lambda_\sharp, e^{\ri \xi_\sharp}\right)\,,\, \lambda_\sharp\right),
\label{F47}\ee
i.e., every admissible triple is given by Lemma 5.5.

It remains to establish the
symplectomorphism between  $\cM_\red^+$ in (\ref{F43}) and $\cD_+ \times \T^n$.
Before going into this, we need some preparation.
We first note $\cM^+_\red$ is open subset of $\cM_\red$ since
$\cD_+$ is an open subset of $\R^n$ and $\cL\colon \cM_0 \to \R^n$
defined in (\ref{C30}) is a continuous, gauge invariant map,
which descends to a continuous map from $\cM_\red$ to $\R^n$.
As a consequence of (\ref{F35}), $\cM_\red^+$ is dense in $\cM_\red$.
It is also true that $\cL$  is an analytic map,
because its components are logarithms of eigenvalues of $gg^\dagger$, and
 (\ref{F36}) ensures that
the eigenvalues of $gg^\dagger$ are pairwise distinct positive numbers for any $g\in \cM_0$.
Let us define $\cM_0^+:= \cL^{-1}(\cD_+)$, and introduce also
$\cM_1^+:= \cM_1\cap \cM_0^+$, as well as the subset $\cN^+\subset \cN$ consisting  of the admissible triples
$(\tilde w, Q, \lambda)$ for which $\lambda \in\cD_+$.
Finally, let $\cS^+\subset \cN^+$ stand for the set of admissible triples parametrized by $\cD_+ \times \T^n$
using (\ref{F40}) with $\lambda\in \cD_+$ and
the phases $e^{\ri \xi_a}$ of $\tilde w_a$ satisfying (\ref{F44}).

Any admissible triple $(\tilde w, Q, \lambda) \in \cN^+$
 is gauge equivalent to a unique admissible triple in $\cS^+$, parametrized
 by $(\lambda, e^{\ri \theta})\in \cD_+ \times \T^n$ with
 \be
  e^{\ri \theta_j } = \frac{\tilde w_j \tilde w_{j+n}^*}{  \vert \tilde w_j \tilde w_{n+j}\vert},
 \qquad j=1,\dots, n.
\label{F50}\ee
 By this  formula,  we can view $e^{\ri \theta}$ as a gauge invariant
function on $\cN^+$, and we also obtain the identification $\cN^+/\T_n \simeq \cS^+$
with respect to the gauge action in (\ref{C29}).
Now we define a map
\be
\psi_+\colon \cM_0^+ \to \cD_+ \times \T^n\equiv \cS^+
\label{psi+}\ee
by composing  a gauge transformation
 $f_0\colon \cM_0^+ \to \cM_1^+$  with the map $\pi_1\colon \cM_1^+ \to \cN^+$
 given by equation (\ref{C13}), and with the map $\cN^+ \to \cS^+$ operating according to (\ref{F50}).
 (The notations are borrowed from Figure 2. See also Remark 3.1.)
Since the $\lambda$-values belonging to $\cD_+$ are regular,
the map $\psi_+$ is smooth (even analytic). It is obviously gauge invariant, surjective and  maps
different gauge orbits to different points.
Therefore $\psi_+$  descends to a one-to-one smooth map
$\Psi_+\colon \cM_\red^+ \to \cD_+ \times \T^n$.
 It was shown in \cite{FM} (without explicitly specifying the domain $\cD_+$ in the calculation)
that $\Psi_+$ satisfies
\be
 \Psi_+^* (\sum_{j=1}^n d \theta_j \wedge d \lambda_j)= \omega_\red^+
\label{F49}\ee
with the restriction $\omega_\red^+$ of the reduced symplectic form on $\cM_\red^+ \subset \cM_\red$.
In particular, the Jacobian determinant of $\Psi_+$ is everywhere non-degenerate, and therefore the inverse map is
also smooth (and analytic).
 \hfill $\square$
\medskip

We finish this section  with a few remarks.
 The strong regularity condition was employed to ensure
that we never divide by zero in the course of the analysis.
 The non-vanishing of $ p_1$ (\ref{F1}) and the first factor of $ p_2$ (\ref{F2}) prevents zero
 denominators in (\ref{F8}), (\ref{F9}) and (\ref{F14}). The non-vanishing of the second factor of
 $ p_2$
 was used in the argument below (\ref{F11}). The last two factors of $ p_2$
 exclude the vanishing of the functions $\cF_k$ (\ref{F15}) or of a component of $\rho$ (\ref{C9}),
 which are not differentiable at those excluded values of $\lambda$ on account of some square roots
 becoming zero.

 Notice from (\ref{F50}) that (because of vanishing denominators) the
 variables $e^{\ri \theta_j}$ cannot all be well-defined at such points where $\lambda$ belongs to the boundary of $\cD$.

Up to this point in the paper,  we have not
used the assumption  (\ref{I13}).
 We shall utilize it  in the following section,
where  we introduce new
 variables that cover also the part of $\cM_\red$  associated with the boundary of $\cD$.
Imposing $\vert u\vert  >\vert v \vert $
ensures, by virtue of $\cD= \overline{\cD_+}$ (\ref{F37}), that the regularity condition (\ref{P12}) holds globally,
since $\lambda_n > \vert v \vert$ is equivalent to $\beta_n >0$.
This in turn ensures, by the arguments developed in Section 3 and Section 4 (see
(\ref{P25}) and (\ref{C26})), that we
have the identification
\be \cM_\red =(K_+(\hat w)\times \T_1)\backslash \cM_1/ \T_{n-1} = \cN/\T_n.
\label{F38}\ee
If $\vert v \vert > \vert u \vert$, then $\beta_n=0$ corresponding to
$\lambda_n = \vert v \vert$ is allowed for elements of $\cM_1$.
As  mentioned after equation (\ref{P29}), this would complicate
the arguments. Also, if $\beta_n =0$, then the corresponding isotropy group
$K_+(\lambda)$ that appears in (\ref{C26}) is larger then $\T_{n-1}$  in (\ref{P13}).
The desire to avoid these complications, together with the symmetry mentioned above (\ref{I13}),
 motivates adopting this assumption in Section 6.

Finally, we recall from \cite{FM} that
\emph{the reduction of  $\cH_1$ (\ref{T13})
gives the RSvD  type Hamiltonian (\ref{I11}) in terms of the Darboux variables $(\lambda, e^{\ri \theta})$}.

\section{The global model $M$ of $\cM_\red$ and  consequences}
\label{sec:6}

We  construct the global model $M$ by bringing every admissible triple $(\tilde w, Q, \lambda)\in \cN$
to a convenient normal form. We then present consequences
for our pair of integrable systems.

\subsection{Construction of the model $M$ of $\cM_\red$}

Adopting the assumption (\ref{I13}), we start with the observation that most (but not all) functions
$\vert \tilde w_a \vert(\lambda)$ contain a factor of the form
\be
\sqrt{\lambda_j - \lambda_{j+1} - \mu}, \quad j=1,\dots, n-1,
\qquad
\sqrt{\lambda_n - \vert u \vert},
\label{S1}\ee
multiplied by a function of $\lambda$ which is strictly positive and analytic
in an open neighbourhood of $\cD\equiv \cD(u,v,\mu)$.
On account of the formula (\ref{F40}), the moduli of the components of $Q$ depend only on $\lambda$,
and for certain indices they are strictly positive, analytic functions.
The precise way in which this happens depends on the sign of $u$, and
now we assume for concreteness that
\be
 \vert u \vert >  \vert v \vert \quad\hbox{and}\quad  u < 0.
\label{S2}\ee
We shall comment on the modifications necessary when this does not hold.

\medskip \noindent
{\bf Lemma 6.1.}
\emph{Under the assumptions (\ref{S2}), for every admissible triple $(\tilde w, Q, \lambda)\in \cN$ we have
\be\ba
&\vert \tilde w_1 \vert = f_1(\lambda),
\\
&\vert \tilde w_j \vert =  \sqrt{\lambda_{j-1} - \lambda_j -\mu} \,f_j(\lambda), \quad j=2,\dots, n-1,\\
&\vert \tilde w_n \vert =  \sqrt{\lambda_n - \vert u \vert}\,\sqrt{\lambda_{n-1} - \lambda_n - \mu}\, f_n(\lambda),\\
&\vert \tilde w_{n+j} \vert =  \sqrt{\lambda_{j} - \lambda_{j+1} -\mu} \,f_{n+j}(\lambda), \quad  j=1,\dots, n-1,\\
&\vert \tilde w_{2n} \vert = f_{2n}(\lambda), \\
\ea\label{S3}\ee
and
\be
\ba
&\vert Q_{j+1, n+j}\vert  = f_{j+1, n+j}(\lambda),\quad j=1,\dots, n-2,\\
&\vert Q_{n, 2n-1}\vert  = \sqrt{\lambda_n - \vert u \vert}\,f_{n, 2n-1}(\lambda),
\ea
\label{S4}\ee
where the $f_i$ and the $f_{j+1, n+j}$ are strictly positive, analytic function in a neighbourhood of $\cD$.
All components of $\rho(\lambda)$ (\ref{C9}) are also analytic functions in a neighbourhood of $\cD$. }
\medskip

It is straightforward  to write explicit formulae for the functions $f_i$ and $f_{j+1,n+j}$.
We shall not use them, but for completeness present some of them in Appendix A.
Here, we note only that, as was pointed out in the proof of Lemma 5.5, the vanishing denominators
of $C_{j+1,n+j}$ in
$Q_{j+1, n+j}$ are cancelled by a zero of $\tilde w_{j+1} \tilde w_{n+j}^*$, for any $j$.
Analogous formulae can be written for all matrix elements of $Q$.
The only  non-displayed matrix element of $Q$ that never vanishes is $Q_{1,2n}$.

The factors (\ref{S1}) lose their smoothness when they become zero, which happens at the boundary of $\cD$.
This is  analogous to the failure of the function $f\colon \C \to \R$ given by $f(z) = \vert z\vert$
to be  differentiable at the origin in $\C$.
Our globally valid new variables will be $n$ complex numbers running over $\C$,  whose
moduli are the factors (\ref{S1}).
Before presenting this, let us  remark that
in terms of a complex variable  the standard symplectic form on $\R^2\simeq \C$
can be written (up to a constant) as $\ri dz \wedge d  z^*$, and the equality
\be
\ri dz \wedge d  z^*=  d r^2 \wedge d\phi \quad\hbox{with}\quad z = r e^{\ri \phi}
\label{S5}\ee
 holds on $\C^*= \C\setminus\{0\}$.
This may motivate one to introduce new Darboux coordinates on $\cD_+ \times \T^n$  like in the next lemma.

\medskip\noindent
{\bf Lemma 6.2.}
\emph{
The following formulae define a diffeomorphism from $\cD_+\times\T^n$ to $(\C^*)^n$
\be
\zeta_j:= \sqrt{\lambda_j - \lambda_{j+1} - \mu} \prod_{l=1}^j
 e^{-\ri \theta_l} \quad\hbox{for}\quad j=1,\dots, n-1,
\quad
\zeta_n:= \sqrt{\lambda_n - \vert u \vert} \prod_{l=1}^n e^{-\ri \theta_l}.
\label{S6}\ee
The symplectic form  that appears in (\ref{F49}) satisfies
\be
\sum_{j=1}^n d\theta_j \wedge d\lambda_j = \ri \sum_{j=1}^n d\zeta_j \wedge d  \zeta_j^*.
\label{S7}\ee
}

Extending the definition (\ref{S6}) to $\cD \times \T^n$,
the boundary of $\cD$ corresponds to the subset of $\C^n$  on which
$\prod_{i=1}^n \zeta_i=0$.
Since we know that the boundary of $\cD$ is part of the admissible $\lambda$ values,
it is already rather clear that $\zeta_i$ as defined above extend to global
coordinates on $\cM_\red$. Nevertheless, this requires a proof.
The proof will enlighten the origin of the complex variables $\zeta_i$.

It is clear from Lemma 6.1 that for any  $(\tilde w, Q, \lambda)\in \cN$ there exists a
unique gauge transformation\footnote{One also sees from this that the action of $\T_n$ on $\cN$ is free.
This can be used to confirm that the effective gauge group (\ref{T34}) acts freely on $\cM_0$.}
 (\ref{C29}) by $\tau=\tau(\tilde w, Q, \lambda)\in \T_n$ (\ref{P15})
such that for the gauge
transformed triple the first and last components of $\tau \tilde w$ are  real and positive and
the components $(\tau Q \tau^{-1})_{j+1, j+n}$ are real and negative for all $j=2,\dots, n-2$.
(The choice of negative sign stems from (\ref{F41}).)
This map can be calculated explicitly.
By using this, we are  able to obtain an analytic, gauge invariant map from $\cM_0$ onto
$\C^n$, which gives rise to a symplectomorphism between $\cM_\red$ and $\C^n$.
Below, we elaborate this statement.

\medskip
\noindent
{\bf Definition 6.3.}  Let $S \subset \cN$ be the set of admissible triples, denoted $(\tilde w^S, Q^S, \lambda)$,
satisfying the following gauge fixing conditions:
\be
\tilde w^S_1 >0, \quad \tilde w^S_{2n}>0, \quad Q^S_{j+1, n+j} <0  \quad\hbox{for}\quad j=1,\dots, n-2.
\label{S8}\ee
As in the proof Theorem 5.6, let
 $\cS^+\subset \cN^+$  denote the set of admissible triples parametrized by $\cD_+ \times \T^n$
using (\ref{F40}) with $\lambda\in \cD_+$ and
the phases $e^{\ri \xi_a}$ of $\tilde w_a$ satisfying (\ref{F44}).
\medskip

We know that $\cS^+$ defines a unique normal form for the elements of $\cN^+\subset \cN$, and
$S$ defines a unique normal form for the whole of $\cN$.
For any $(\tilde w, Q, \lambda)\in \cN$, we  define the $n$ phases $X_1, X_n, X_{j+1,n+j}\in \U(1)$ by
writing
\be
\tilde w_1 = X_1 f_1(\lambda),\quad  \tilde w_{2n} = X_{2n} f_{2n}(\lambda),\quad
Q_{j+1, n+j} = - X_{j+1, n+j} f_{j+1, n+j}(\lambda)
\label{S9}\ee
for every  $j=1,\dots, n-2$.  The map $(\tilde w, Q, \lambda)\mapsto (\tilde w^S, Q^S,\lambda)$
sends any admissible triple  to the intersection of its $\T_n$ orbit (defined by (\ref{C29}))
with $S$, which is given by
\be
(\tilde w^S, Q^S,\lambda) =( \tau \tilde w, \tau Q \tau^{-1},\lambda )
\,\,\, \hbox{with}\,\,\,
 \tau_1 = X_1^{-1},\quad \tau_{2n}= X_{2n}^{-1},\quad \tau_j =X_1^{-1}\prod_{i=1}^{j-1} X_{i+1,n+i}^{-1}
\label{S10} \ee
 for  $j=2,\dots, n-1$. This yields $\tilde w^S$ and $Q^S$
 as gauge invariant functions on $\cN$, and by using them we can define the
 $\C^n$ valued gauge invariant map $\pi_\cN\colon (\tilde w, Q, \lambda) \mapsto \zeta$  on $\cN$ as follows:
   \be \ba
   &\zeta_j(\tilde w, Q, \lambda) := \tilde w^S_{n+j}/f_{n+j}(\lambda), \quad j=1,\dots, n-1,\\
&\zeta_n(\tilde w, Q, \lambda):= (Q^S_{n, 2n-1})^*/f_{n,2n -1}(\lambda).
\ea\label{S11}\ee
For the remaining components of the function $\tilde w^S$ given by (\ref{S10}), we find
\be
\tilde w^S_j = \zeta_{j-1} f_j(\lambda), \quad j=2,\dots, n-1,\quad \tilde w^S_n=
\zeta_n^* \zeta_{n-1} f_n(\lambda)
\label{S12}\ee
with the functions of $\lambda$ in (\ref{S3}), and of course $\tilde w^S_1 = f_1(\lambda)$ and
 $\tilde w^S_{2n} = f_{2n}(\lambda)$.
The function $Q^S$ (\ref{S10}) is given by substituting $\tilde w^S$ for $\tilde w$ in the formula (\ref{F40}).

Equation (\ref{S12}) can be checked
by writing every $(\tilde w, Q, \lambda)$  in terms of
 $(\lambda, e^{\ri \xi}) \in \cD \times \T^{2n}$ as in (\ref{F40}), cf.~Lemma 5.5.
By applying this, we obtain, for $j=1,\dots, n-1$,
\be
\zeta_j = \sqrt{ \lambda_j - \lambda_{j+1} -\mu} \prod_{l=1}^j  e^{-\ri \xi_l} e^{\ri \xi_{n+l}}
\quad\hbox{and}\quad
 \zeta_n = \sqrt{\lambda_n - \vert u \vert} \prod_{l=1}^n  e^{-\ri \xi_l} e^{\ri \xi_{n+l}}.
\label{S13}\ee
This shows manifestly that the range of $\zeta$ covers the whole of $\C^n$.
If we restrict this formula to $\cS^+$,  parametrized by $\cD_+ \times \T^n$ using (\ref{F44}),
\emph{then we recover  our previous formulae (\ref{S6})}.
We now  summarize  these claims.

\medskip \noindent
{\bf Proposition 6.4.} \emph{The $\T_n$ gauge invariant map
$ \pi_\cN\colon (\tilde w, Q, \lambda) \mapsto \zeta$ exhibited
in (\ref{S11})
induces a bijection between $\cN/\T_n$
and $\C^{n}$.  The restriction of the component functions $\zeta_i$ to $\cS^+ \subset \cN$
is given by the formula (\ref{S6}).
The inverse map from $\C^n$ to $S\simeq \cN/\T^n$
can be written down explicitly by first expressing $\lambda$ in terms of $\zeta$ as
\be
\lambda_j = \vert u \vert + (n-j) \mu + \sum_{l=j}^n \vert \zeta_l \vert^2, \quad j=1,\dots, n,
\label{S14}\ee
then expressing $\tilde w^S$ by means of $\zeta$ using  (\ref{S11}) and (\ref{S12}), and finally obtaining
$Q^S$ as a function of $\zeta$ via substitution of  $\tilde w^S(\zeta)$ for $\tilde w$
in the formula (\ref{F40}).}

 \medskip
 \noindent
 {\bf Proof.} The surjectivity onto $\C^n$ was explained above, and
 the injectivity is clear because we can explicitly write down the inverse from $\C^n$ onto
 the global cross-section $S$ of the $\T_n$ action on $\cN$.
 \hfill $\square$

\medskip

Our main theorem says that the construction just presented gives a global
model of $\cM_\red$:
\be
(M, \omega) \equiv (\C^n, \omega_{\can})
\quad \hbox{with}\quad
\omega_{\can} = \ri \sum_{j=1}^n d \zeta_j \wedge d  \zeta_j^*.
\label{S15}\ee

\medskip
\noindent
{\bf Theorem 6.5.}  \emph{Take an arbitrary element $g_0\in \cM_0$ and pick $g(g_0)$ to be an element of $\cM_1$
which is gauge equivalent to $g_0$.
Then define the map $\psi\colon \cM_0 \to \C^n$ by the rule
\be
\psi\colon g_0 \mapsto \zeta\left(\tilde w(g(g_0)), Q(g(g_0)), \lambda(g(g_0))\right),
\label{S16}\ee
combining (\ref{S11}) with the map $ \cM_1\ni g \mapsto (\tilde w, Q, \lambda)\in \cN$ given by
equations  (\ref{C12}) and (\ref{C13}).
The map $\psi$ is analytic, gauge invariant  and it descends to a diffeomorphism $\Psi\colon \cM_\red \to \C^n$
having the symplectic property
\be
\Psi^* (\omega_{\can}) = \omega_\red.
\label{S17}\ee
}

\medskip
\noindent
{\bf Proof.}
Since it does not depend on the choice for $g(g_0)$,
the analyticity  of $\psi$ follows from the possibility of an analytic local choice (see Remark 3.1) and the explicit formulae involved in the definition
(\ref{S16}).
Its bijective character is a direct consequence of Proposition 6.4.
The symplectic property  follows from Theorem 5.6
 and a density argument.
 Namely, on $\cM_\red^+$ we can convert $\Psi_+$ satisfying (\ref{F49}) into $\Psi$
 by means of the map $(\lambda, e^{\ri \theta}) \mapsto \zeta$ as given  by (\ref{S6}).
 This and Lemma 6.2 imply the equality (\ref{S17}) for the restriction of $\Psi$ on $\cM_\red^+$,
  and then the equality
 extends to the whole space by the smoothness of $\Psi$, $\omega_\can$ and $\omega_\red$.
 As a consequence of (\ref{S17}),  the inverse map is
smooth as well.
\hfill $\square$

\medskip\noindent
{\bf Remark 6.6.}
The formulae of the complex variables used in Section 2.2 can be converted into those applied
in this section by introducing new  `tilded variables' as
\be
\tilde \lambda_j := - \hat\lambda_{n+1-j} + c,\quad
\tilde \theta_j := - \hat \theta_{n+1-j},\quad
\tilde \cZ_{k} := \cZ_{n-k},\quad  \tilde\cZ_n = \cZ_n,
\label{tildedvars}\ee
for $j=1,\dots, n$ and $k=1,\dots, n-1$.
Then $\tilde \cZ$ depends on $\tilde \lambda, \tilde \theta$
by the same formula (\ref{S6}) whereby $\zeta$ depends on $\lambda, \theta$.
By choosing the constant $c$ appropriately, the domain of $\tilde \lambda$ also becomes identical
to the domain of $\lambda$.

\medskip\noindent
{\bf Remark 6.7.}
As promised,  we now comment on the modification of the construction for the cases when (\ref{S2}) does not hold.
If instead we have $\vert u \vert > \vert v \vert$ and $u>0$, then the definition (\ref{S6}) is
still applicable, but (\ref{F15}) implies that
the factor $\sqrt{\lambda_n- \vert u \vert}$ is contained in $\vert \tilde w_{2n}\vert$
instead of $\vert\tilde w_n\vert$,
and  thus $\vert Q_{n,2n-1}\vert$ does not contain this factor
 (cf.~(\ref{S3})). Then one may proceed by defining  a global cross-section $S \subset \cN$ with the
help of the gauge fixing conditions $\tilde w_1^S>0$ and $Q^S_{j+1, n+j}<0$ for all
 $j=1,\dots, n-1$ (cf.~(\ref{S8})).
The construction works quite similarly to the above one, and all consequences described in
the next subsection remain true.
As was discussed in the Introduction, we can impose (\ref{I13}) without loss of generality.
Nevertheless, it could  be a good exercise  to detail the construction of
the counterpart of our model $M$   when (\ref{I13}) does not hold. We only note that one must then define
$\zeta_n$ in such a way that $\vert \zeta_n\vert = \sqrt{\lambda_n - \vert v \vert}$  and use that,
on account of (\ref{C5}),
this factor is contained in a matrix element of $\rho(\lambda)$ (\ref{C9}).

  \subsection{Consequences of the model of $M$ and the duality map}

Our symplectic reduction yields two Abelian Poisson algebras,
 $\fH_\red^1$ and $\fH^2_\red$, on the reduced phase space $(\cM_\red, \omega_\red)$.
Concretely, $\{ \cH_j^\red\}_{j=1}^n $, descending from the functions $\cH_j$ (\ref{T13}), is a generating set for $\fH^1_\red$
and $\{ \hat \cH_j^\red\}_{j=1}^n$,  descending from the  functions $\hat \cH_j$ (\ref{T9}), is a generating set for
$\fH^2_\red$.
We have two models $(\hat M, \hat \omega)$ and $(M,\omega)$
of $(\cM_\red, \omega_\red)$, endowed with the symplectomorphisms
\be
\hat \Psi\colon \cM_\red \to \hat M, \qquad \Psi\colon \cM_\red \to M,
\label{S18}\ee
and the duality map
\be
\quad \cR:= \hat \Psi \circ \Psi^{-1}\colon M \to \hat M.
\label{S19}\ee
The restriction of
\be
\hat H\equiv \hat \cH_1^\red \circ \hat \Psi^{-1}
\label{hatHid}\ee
 to $\hat M^o = (\C^*)^n $ acquires the
form
(\ref{I6}) if $\hat M^o$ is parametrized by $\widehat\cD_+ \times \T^n$ as described in Section 2.2,
and the restriction of
\be
H \equiv  \cH_1^\red \circ \Psi^{-1}
\label{Hid}\ee
 to $M^o = (\C^*)^n$ takes
the form (\ref{I11}) if  $M^o = (\C^*)^n$ is parametrized by $\cD_+ \times \T^n$ as
given by (\ref{S6}).
The interpretations of the  reduced Hamiltonians  from the perspective of the model
$(\hat M,\hat \omega)$ were
outlined
in Section 2.2,
and we now discuss the significance of the model $(M, \omega)$.

The first basic point about $M$ is that the flow of the RSvD type Hamiltonian
$H$ (\ref{I11}) is not complete on the dense open subset $M^o \subset M$, while its
reduction origin ensures completeness on $M$.
The flows of all  $\cH_j^\red \circ \Psi^{-1}$ are also complete on $M$, simply since they are projections
of complete flows on the unreduced phase space $\cM$.
The second basic point is that $(M,\omega)$ serves naturally as action-angle phase space
for the integrable Hamiltonians $\hat \cH_j^\red \circ {\hat \Psi}^{-1}$, which include the RSvD type Hamiltonian
(\ref{I6}).
Indeed,   the map  $\cR$ `trivializes'
the  Hamiltonians $\hat \cH_j^\red \circ \hat \Psi^{-1}$, since we have
\be
(\hat \cH_j^\red \circ \hat \Psi^{-1})\circ \cR = \hat \cH_j^\red \circ \Psi^{-1}= \sum_{l=1}^n \cosh (2 j \lambda_l).
\label{S20}\ee
Thus, the functions $\lambda_l\colon M\to \R$ are action variables for the (completed)
integrable many-body system (\ref{I6}) on $\hat M$.
The actions $\lambda_l$  are related by a $\GL(n,\Z)$ transformation combined with a constant shift
to the distinguished
action variables defined by the functions $\vert \zeta_i \vert^2$ on $M = \C^n$.
These latter action variables generate the standard $\T^n$ action on
$M= \C^n$.
The origin $\zeta=0$ is a fixed point for the torus action, and it represents
the unique joint minimum of the Hamiltonians (\ref{S20}).
Moreover, this is the only equilibrium point that any single Hamiltonian of the form (\ref{S20}) possesses.

It follows from the above that $ \cR(0)\in \hat M $ is a joint equilibrium point for the Hamiltonians
$\hat \cH_j^\red \circ \hat \Psi^{-1}$.
It also follows that each Hamiltonian  $\hat \cH_j^\red\circ \hat \Psi^{-1}$
is non-degenerate (has no extra conserved quantities), because this property of the equivalent Hamiltonians
(\ref{S20})
is easily seen.
 Of course, one can write down the analogues of equations (\ref{T49}) -- (\ref{T52}) for the
flows of the Hamiltonians (\ref{S20}) on $M$.
For any fixed $j$, the counterparts of the $n$ frequencies  (\ref{T51}) are
given by $\Omega_{j,a}(\lambda) = 2 j \sinh{(2j\lambda_a)}$, which
generically are independent over the field of
rational numbers.
The existence of an equilibrium point for $\hat H$ (\ref{I6})
is not obvious. It is  an open problem to find the $\cZ$-coordinates of $\cR(0)\in \hat M$; we believe that it
lies inside the dense open set $\hat M^o$.
A similar  open problem is to find $\cR^{-1}(0)\in M$, which gives the unique joint equilibrium for the
Hamiltonians $\cH_j^\red \circ \Psi^{-1}$.

We have established the alternative
interpretations of the $\vert \zeta_i \vert^2\in C^\infty(M)$ as action variables  for
$\hat \cH_j^\red \circ \hat \Psi^{-1}$ and
global position variables   $\cH_j^\red \circ \Psi^{-1}$, respectively.
At the same time, the functions $\vert \cZ_i \vert^2\in C^\infty(\hat M)$ serve as actions for
$\cH_j^\red \circ \Psi^{-1}$
and global position variables for $\hat \cH_j^\red \circ \hat \Psi^{-1}$.
This shows  that the integrable many-body systems engendered by the `main Hamiltonians' displayed in
(\ref{I6}) and (\ref{I11}) are indeed in action-angle duality.

A special feature of the dual pair at hand is that the action-angle phase spaces $(M,\omega)$ and
$(\hat M,\hat \omega)$
are also \emph{the same} in an obvious manner, namely, both are equal to $(\C^n,\omega_\can)$.
 Distinguished action variables of both systems generate the standard
torus action on $\C^n\simeq \R^{2n}$ equipped with its canonical symplectic form.
It is by no means true that every Liouville integrable system corresponds to a globally well-defined
Hamiltonian torus action, and for global torus actions there could be several
inequivalent possibilities.
Integrable many-body systems in action-angle duality live on symplectomorphic phase spaces,
but their respective action variables cannot in general be intertwined by a symplectomorphism.
Apart from the current example and self-dual systems, such an action-intertwining symplectomorphism
was previously found only for dual pairs of purely scattering systems, such as
the hyperbolic Sutherland system and its Ruijsenaars dual  \cite{SR88}, and the analogous
$\mathrm{BC}_n$ systems \cite{P}.

It may be worth stressing that  the duality map $\cR$ (\ref{S19}) is just the identity map on $\cM_\red$
written in terms of two distinct models. On the other hand, the
map $M \to \hat M$ given by $\zeta \mapsto \cZ = \zeta$
encodes
a  non-trivial map on $\cM_\red$, for which $\Psi^{-1}(\zeta) \mapsto \hat \Psi^{-1}(\zeta)$,
$\forall \zeta\in \C^n$.

We end by remarking that one can perform semiclassical quantization for both systems using their respective
action variables.  Even more,  one can quantize any action variable of the form
$\vert \zeta_j\vert^2\in C^\infty(\C)$ by the replacement
\be
 \zeta_j^*  \zeta_j  \longrightarrow {\hat \zeta_j}^\dagger \hat \zeta_j,
\label{S21}\ee
where the hatted letters stand for annihilation and creation operators on the standard Fock space.
In this manner, one obtains that the spectrum of each action variable $\vert \zeta_j \vert^2$ consists
of all non-negative integers.  This then gives immediately  the (semi-classical) spectra of the corresponding
integrable Hamiltonians. Regarding the Hamiltonians  (\ref{S20}),
 one simply expresses $\{\lambda_i\}$ in terns of $\{\vert \zeta_j \vert^2\}$.
One can deal with the Hamiltonians $\cH_j^\red\circ \hat \Psi^{-1}$ (\ref{T43})
in the same spirit.

\section{Discussion and outlook}
\label{sec:7}

We have  presented
a thorough description of
the models $M$ and $\hat M$ of the reduced phase space $\cM_\red$ (\ref{T33})  and gained
a detailed understanding of how these models are equipped with a pair of
 integrable many-body systems in action-angle duality.
Our principal result is that we have established the validity
of Figure 1 of the Introduction for the case at hand.
In particular, we have seen that $\lambda\colon M \to \R^n$  yields via the duality map $\cR$
the  momentum map
 for the torus action associated with the integrable Hamiltonians
 $\hat \cH_j^\red\circ \hat \Psi^{-1}$ that contain $\hat H$ (\ref{I6})
 and at the same time it provides global position variables for the
 Hamiltonians $\cH^\red_j\circ \Psi^{-1}$ that contain $H$ (\ref{I11}).
 This and the analogous dual interpretations for
 the map $\hat \lambda\colon \hat M \to \R^n$ are  explained in Section 2.2 and Section 6.2.
 To put it slightly differently, we have seen that
  $\{ \lambda_j\}$ and $\{\hat \cH_j^\red\circ \Psi^{-1}\}$ (\ref{S20})
 are alternative generating sets for the Abelian Poisson algebra $\fP$ on $(M,\omega)$, while
 $\{ \hat \lambda_j\}$ and $\{ \cH_j^\red \circ \hat \Psi^{-1}\}$ (\ref{T44}) provide alternative
 generating sets for $\hat \fP$ on $(\hat M, \hat \omega)$.

The main technical achievement of this paper is the construction of the model $M$,
which is summarized by Figure 2  and  Theorem 6.5.
The constructions of the maps $\psi$ and $\hat \psi$ that feature in the two figures rely
 respectively on the singular value decomposition  and on the generalized Cartan decomposition
of certain matrices, and other algebraic operations.
These maps, and especially
the duality map $\cR$, cannot be presented explicitly, basically since
the eigenvalues of higher rank matrices cannot be given in closed form.
Nevertheless,  the duality proves very
useful for understanding the qualitative features of the  respective systems.

Our study gives rise to the first  example of systems in  duality for which
the two systems are different (not a self-dual case)  and both have quasi-periodic motions
on compact Liouville tori.
The duality map $\cR$  allowed us to demonstrate
that in our case each one of the two systems $(M,\omega, \fH,\fP,H)$ and
$(\hat M, \hat\omega, \hat\fH,\hat\fP,\hat H)$
 has a unique equilibrium position, which corresponds to the origin
in $\C^n$ used to represent  both $M$ and $\hat M$.
We also pointed out that each reduced Hamiltonian $\cH_j^\red$ and $\hat \cH_j^\red$ possesses
Abelian commutants in
the Poisson algebra $C^\infty(\cM_\red)$.
As another spin-off, let us now explain  that
the particle positions evaluated along any  fixed phase space trajectory of our Hamiltonians
stay in a compact set, i.e., all motions are bounded.
Indeed, any trajectory of $\cH_j^\red \circ \Psi^{-1}$ is contained in a set
 $(\hat\lambda \circ \cR)^{-1}(\hat\lambda_0)$ for some
$\hat\lambda_0 \in \R^n$, which is  compact, since---being  equivalent to the standard
$\T^n$ momentum map on $(\C^n,\omega_\can)$ (\ref{T40})---the map $\hat\lambda \colon \hat M\to \R^n$ is  proper.
This compact subset of $M$ is sent by $\lambda$  onto a compact subset of  $\R^n$,
simply because $\lambda\colon M\to \R^n$ is continuous.
A similar argument can be applied to the trajectories generated by
 the Hamiltonians $\hat \cH_j^\red \circ \hat \Psi^{-1}$ as well.

We remark that in principle we can derive Lax pairs
for our systems, since we know the `unreduced Lax matrices' (see (\ref{T9}) and (\ref{T13})) that
generate the Abelian Poisson algebras $\fH^1$ and $\fH^2$ on $\cM$, and
those unreduced Lax matrices satisfy Lax equations already before reduction \cite{FM,M1}.
The specific formulae should be worked out and compared with the Lax matrices obtained
recently in \cite{PG}.

We have seen that the  complex `oscillator variables' provide an easy way for finding
the semiclassical spectra of the actions, by (\ref{S21}), and thus also the  spectra of
the many-body Hamiltonians.
It is an interesting problem for future work to compare
 this   `action-angle quantization' with a `Schr\"odinger quantization'
of the RSvD type many-body Hamiltonians (\ref{I6}) and (\ref{I11})  built on analytic difference operators.
For this, the recent paper by van Diejen and Emsiz \cite{vDE}  should serve
 as  a good starting point.

 Another promising project is to explore reductions of the Heisenberg double of $\SU(2n)$
at generic values of the momentum map. This is expected to produce extensions with internal degrees of freedom of  the many-body
systems (\ref{I6}) and (\ref{I11}). A suitably generalized version of action-angle duality
should hold also for such systems, analogously to the systems
investigated  by Reshetikhin  \cite{Resh1,Resh2}.

Finally, we wish to draw attention to our supplementary new result presented in Appendix B, where we
show how the Hamiltonian $H$ (\ref{I11}) can be recovered as a scaling limit of
van Diejen's 5-parametric integrable Hamiltonians \cite{vD1}.
We stress that our reduced
  Hamiltonians automatically have complete flows on $\cM_\red$,  while
the completeness of the flow for general real forms of van Diejen's systems has not
yet been studied. However, see \cite{PG}, and also
\cite{PR} for a detailed study of classical scattering in a 2-parameter hyperbolic  case.
The most intriguing open problem in this area  is to find a
 Hamiltonian reduction treatment  for van Diejen's   5-parametric systems.
This would  enhance their group theoretic understanding, and
would also help to explore their classical dynamics.

\bigskip\bigskip \noindent\bf Acknowledgements. \rm
We wish to thank Alexei Rosly and Simon Ruijsenaars for useful
discussions.
We are also grateful to Tam\'as G\"orbe and  G\'abor Pusztai for comments on the manuscript.
L.F.~is indebted to Youjin Zhang for hospitality at Tsinghua University during the
final stage of the work.
This work was supported in part by the Hungarian Scientific Research
Fund (OTKA) under the grant K-111697.

\renewcommand{\thesection}{\Alph{section}}
\setcounter{section}{0}
\section{Some explicit formulae}

In this appendix we display the explicit formulae of some of the functions that appear in Lemma 6.1.
We begin by noting that
$f_1(\lambda) = \sqrt{\cF_1(\lambda)}$ and $f_{2n}(\lambda) = \sqrt{\cF_{2n}(\lambda)}$,
since for these suffixes the functions (\ref{*}) are positive in a neighbourhood of the domain
$\cD=\overline{\cD_+}$ (\ref{F37}). We here used the assumption (\ref{S2}) and the explicit
formula (\ref{F15}).
 To deal with  the other components $\vert \tilde w_j\vert$ in (\ref{S3}), we use the analytic function
 \be
 J(x)= \sinh(x)/x,
 \label{A1}\ee
 which is positive for all $x\in \R$. Then we have the following formulae.
 First,
 \be
 \ba
 &f_j(\lambda) = \Bigl[J(\lambda_{j-1} - \lambda_j - \mu)
  \frac{ e^{-\mu}\sinh(\mu)}{\sinh(2\lambda_j)}
  \frac{(e^{2 \lambda_j} - e^{-2u})\sinh(\lambda_j+ \lambda_{j-1} + \mu)}
  {\sinh(\lambda_{j-1} + \lambda_j) \sinh(\lambda_{j-1} - \lambda_j)}
    \Bigr]^{\frac{1}{2}}  \\
\quad &\times \Bigl[
\prod_{\substack{i=1\\(i\neq j,j-1)}}^{n} \left(\frac{\sinh(\lambda_j+\lambda_i+\mu)\sinh(\lambda_j-\lambda_i+\mu)}
{{\sinh(\lambda_j-\lambda_i)}\sinh(\lambda_j+\lambda_i)}\right)
  \Bigr]^{\frac{1}{2}}, \quad  j=2,\dots, n-1,
 \ea
 \label{A2}\ee
 then
 \be
 \ba
 &f_n(\lambda) = \Bigl[ 2 J(\lambda_{n-1} - \lambda_n - \mu)
  \frac{ \sinh(\mu)}{\sinh(2\lambda_n)}
  \frac{e^{ \lambda_n-u -\mu }\sinh(\lambda_n+ \lambda_{n-1} + \mu)}
  {\sinh(\lambda_{n-1} + \lambda_n) \sinh(\lambda_{n-1} - \lambda_n)}
    \Bigr]^{\frac{1}{2}}  \\
\quad &\times \Bigl[J(\lambda_n - \vert u \vert)
\prod_{\substack{i=1\\(i\neq n,n-1)}}^{n} \left(\frac{\sinh(\lambda_n+\lambda_i+\mu)\sinh(\lambda_n-\lambda_i+\mu)}
{{\sinh(\lambda_n-\lambda_i)}\sinh(\lambda_n+\lambda_i)}\right)
  \Bigr]^{\frac{1}{2}},
 \ea
 \label{A3}\ee
 and finally
\be
 \ba
 &f_{n+j}(\lambda) = \Bigl[J(\lambda_{j} - \lambda_{j+1} - \mu)
  \frac{ e^{-\mu}\sinh(\mu)}{\sinh(2\lambda_j)}
  \frac{(e^{-2u}-e^{-2 \lambda_j}  )\sinh(\lambda_j+ \lambda_{j+1} - \mu)}
  {\sinh(\lambda_{j} + \lambda_{j+1}) \sinh(\lambda_{j} - \lambda_{j+1})}
    \Bigr]^{\frac{1}{2}}  \\
\quad &\times \Bigl[
\prod_{\substack{i=1\\(i\neq j,j-1)}}^{n} \left(\frac{\sinh(\lambda_j+\lambda_i-\mu)\sinh(\lambda_j-\lambda_i-\mu)}
{{\sinh(\lambda_j-\lambda_i)}\sinh(\lambda_j+\lambda_i)}\right)
  \Bigr]^{\frac{1}{2}}, \quad  j=1,\dots, n-1.
 \ea
 \label{A4}\ee
 It is easy to see from (\ref{F15}), (\ref{*}) that (\ref{S3}) holds with the above formulae.
 Combining  (\ref{F40}) and (\ref{F41}) with (\ref{S3}), one  can write explicit
  formulae for  the functions in
 (\ref{S4}) as well.
 The main point is that the vanishing denominators $\sinh(\lambda_j - \lambda_{j+1} - \mu)$ of $C_{j+1, n+j}$
 (\ref{F41})  cancel for each $j=1,\dots, n-1$.
The formulae are not enlightening and we omit them.

 \renewcommand{\thesection}{\Alph{section}}
\section{The relation of  $H$ (\ref{I11}) to van Diejen's  Hamiltonian}

 Our starting point is the following  real form
 of van Diejen's  Hamiltonian \cite{vD1}, with real parameters $a,b,c,d,\mu$,
\be\label{hvd}
H_{\rm{vD}}[\mu;a,b,c,d]\,(\lambda,\theta) = \sum_{j=1}^n  (\cos\theta_j) (V_jV_{-j})^{1/2}(\lambda)
\,-\,
\frac12\sum_{j=1}^n(V_j+V_{-j})(\lambda),
\ee
with $V_{\pm j}=V_{\pm j}^{(1)} V_{\pm j}^{(2)}$ and  $V_{\pm j}^{(1,2)}$ given by
\be\label{V12pm}
\ba
V_{\pm j}^{(1)}(\lambda) &= \frac{\cosh(a\pm \lambda_j)\cosh(b\pm \lambda_j)\sinh(c\pm \lambda_j)\sinh(d\pm \lambda_j)}{\cosh^2\lambda_j\sinh^2\lambda_j}\\
V_{\pm j}^{(2)}(\lambda) &=  \prod_{k\neq j}^n\frac{\sinh\bigl(\mu\pm(\lambda_j+\lambda_k)\bigr)\sinh\bigl(\mu\pm(\lambda_j-\lambda_k)\bigr)}{\sinh(\lambda_j+\lambda_k)\sinh(\lambda_j-\lambda_k)}.
\ea
\ee
For convenience, we shall refer to the two terms in the formula for $H_{\mathrm{vD}}(\lambda, \theta)$ as ``kinetic'' and ``potential''.

We will prove the following result.

\begin{proposition}\label{appbprop}
The Hamiltonian in (\ref{I11}) is a special limiting case of the
van Diejen Hamiltonian. Specifically,  on the domain $\cD_+(u,v,\mu) \times \T^n$ (\ref{I9}), we have
\be
H =  e^{v-u}\lim_{\substack{a\rightarrow-\infty\\ b\rightarrow+\infty\\ c\rightarrow u,\,  d\rightarrow v}}\Bigl(4 e^a e^{-b} H_{\rm{vD}}[\mu;a,b,c,d]\Bigr) +n.
\ee
\end{proposition}
Before giving the proof of this result, let us state an intermediate one.

\begin{lemma}\label{appblemma}
The product in the kinetic term can be expressed in the form
\be\label{kvd}
\ba
\bigl(V_jV_{-j}\bigr)(\lambda) &= \left(1+\frac{\sinh^2a}{\cosh^2\lambda_j}\right) \left(1+\frac{\sinh^2b}{\cosh^2\lambda_j}\right) \left(1-\frac{\sinh^2c}{\sinh^2\lambda_j}\right) \left(1-\frac{\sinh^2d}{\sinh^2\lambda_j}\right)\\
&\qquad\qquad\times  \prod_{k\neq j}^n\left(1-\frac{\sinh^2\mu}{\sinh^2(\lambda_j-\lambda_k)}\right)\left(1-\frac{\sinh^2\mu}{\sinh^2(\lambda_j+\lambda_k)}\right),
\ea
\ee
and the potential term in (\ref{hvd}) may be written in the form
\be\label{pvd}
\ba
&- \frac12\sum_{j=1}^n(V_j+V_{-j})(\lambda)
=
\frac{1}{\sinh^2\mu}\cosh(a)\cosh(b)\sinh(c)\sinh(d) \prod_{k=1}^{n}\left(1-\frac{\sinh^2\mu}{\sinh^2\lambda_k}\right)\\
&\qquad +\frac{1}{\sinh^2\mu}\sinh(a)\sinh(b)\cosh(c)\cosh(d)\prod_{k=1}^{n}\left(1+\frac{\sinh^2\mu}{\cosh^2\lambda_k}\right)
+C[\mu;a,b,c,d]
\ea
\ee
with constant
\be\label{cvd}
\ba
C[\mu;a,b,c,d] &=
\frac{1}{2\sinh^2\mu}\Bigl[\cosh(a-b)\cosh(c-d) - \bigl(\cosh(a+b-\mu)\cosh(c+d-\mu) \Bigr]\\
&\qquad
-\frac{\sinh\bigl(a+b+c+d + (2n-1)\mu\bigr)}{2\sinh\mu}.
\ea
\ee
\end{lemma}

\begin{proofof}{Proposition}{appbprop}

Implementing the limit for the potential term, making use of (\ref{pvd}), yields
\be
\ba
&\lim_{\substack{a\rightarrow-\infty\\  b\rightarrow +\infty}} e^ae^{-b}\left(-\frac12\sum_{j=1}^n(V_j+V_{-j})\right)
=\\
&\quad\quad
\frac14\frac{\sinh c\sinh d}{\sinh^2\mu}
\prod_{k=1}^n\left(1 - \frac{\sinh^2\mu}{\sinh^2\lambda_k}\right)
-
\frac14\frac{\cosh c\cosh d}{\sinh^2\mu}
\prod_{k=1}^n\left(1 + \frac{\sinh^2\mu}{\cosh^2\lambda_k}\right)
+
\frac14 \frac{\cosh(c-d)}{\sinh^2\mu}.
\ea
\ee

Applying the same limit to the kinetic term, using (\ref{kvd}), we obtain
\be
\ba
\lim_{\substack{a\rightarrow-\infty\\ b\rightarrow+\infty}}e^{2a}e^{-2b}V_jV_{-j}
&=
\frac1{16}\frac1{\cosh^4\lambda_j}\left(1-\frac{\sinh^2c}{\sinh^2\lambda_j}\right)\left(1-\frac{\sinh^2d}{\sinh^2\lambda_j}\right)\\
&\qquad\times
\prod_{k\neq j}^n\left(1-\frac{\sinh^2\mu}{\sinh^2(\lambda_j+\lambda_k)}\right) \left(1-\frac{\sinh^2\mu}{\sinh^2(\lambda_j-\lambda_k)}\right).
\ea
\ee

Putting these together, we obtain
\be
\ba
&\lim_{\substack{a\rightarrow-\infty\\ b\rightarrow+\infty\\ c\rightarrow u,\, d\rightarrow v}}\Bigl(4e^ae^{-b}H_{\rm{vD}}[\mu;a,b,c,d]\,(\lambda,\theta)\Bigr) = \\
&\sum_{j=1}^n\frac{\cos\theta_j}{\cosh^2\lambda_j}\left[\left(1-\frac{\sinh^2u}{\sinh^2\lambda_j}\right)
\left(1-\frac{\sinh^2v}{\sinh^2\lambda_j}\right)
\prod_{k\neq j}^n\left(1-\frac{\sinh^2\mu}{\sinh^2(\lambda_j+\lambda_k)}\right) \left(1-\frac{\sinh^2\mu}{\sinh^2(\lambda_j-\lambda_k)}\right)\right]^{1/2}\\
&\qquad
 +\frac{\sinh u\sinh v}{\sinh^2\mu}
\prod_{k=1}^n\left(1 - \frac{\sinh^2\mu}{\sinh^2\lambda_k}\right)
-
\frac{\cosh u\cosh v}{\sinh^2\mu}
\prod_{k=1}^n\left(1 + \frac{\sinh^2\mu}{\cosh^2\lambda_k}\right)
+
\frac{\cosh(u-v)}{\sinh^2\mu}.
\ea
\ee
\end{proofof}

\begin{proofof}{Lemma}{appblemma}

Checking (\ref{kvd}) is straightforward. To derive the formula for the potential term, let us define the  meromorphic one-form
\be
\Omega(z) := F(z)dz,
\ee
with the function $F$ defined by
\be
F(z) = \frac12\frac{(Az+A^{-1})(Bz+B^{-1})(Cz-C^{-1})(Dz-D^{-1})}{(\alpha^{-2}-1)z(z^2-1)(z^2-\alpha^2)}\left(\prod_{a=1}^{2n}\frac{\alpha^{-1} z\Lambda_a-\alpha}{z-\Lambda_a}\right).
\ee
The poles of $\Omega(z)$ are at $z=0$, $z=\pm1$, $z=\pm\alpha$, $z=\infty$, $z=\Lambda_a$, and the sum of the residues is zero. Thus we have
\be\label{appbzerosum}
-\sum_{a=1}^{2n}\rez_{z=\Lambda_a}\Omega(z) = \left(\rez_{z=+1}+\rez_{z=-1}+\rez_{z=0}+\rez_{z=\infty}+\rez_{z=+\alpha}+\rez_{z=-\alpha}\right)\Omega(z).
\ee
Upon making the substitutions
\be
\alpha=e^{-\mu},\quad A=e^a,\quad B=e^b,\quad C=e^c,\quad D=e^d,\quad
 \Lambda_j=e^{2\lambda_j},\quad \Lambda_{n+j}= e^{-2\lambda_j},
\ee
(\ref{appbzerosum}) is the same as (\ref{pvd}). That is\\
---the sum of the residues at $z=\Lambda_1,\dots, \Lambda_{2n}$ 
is $(-1)$ times the van Diejen potential,\\
---the sum of the residues at $z=\pm1$ yields the first two terms on the rhs of (\ref{pvd}),\\
---the sum of the residues at $z=\pm\alpha$ yields the first line on the rhs of (\ref{cvd}),\\
---the sum of the residues at $z=0$ and $z=\infty$ yields the second line on the rhs of (\ref{cvd}).

\end{proofof}

 \end{document}